\documentclass[5p, times]{elsarticle}

\usepackage{graphicx}
\usepackage{lineno}
\usepackage{tabularx}
\usepackage{multirow}
\usepackage{multicol}
\usepackage{textcomp}
\usepackage{colortbl}
\usepackage{numprint}
\usepackage{textcomp}
\usepackage{booktabs}
\usepackage{enumitem}
\usepackage{textcomp}
\usepackage{comment}
\usepackage[utf8]{inputenc}
\usepackage{array}
\usepackage{float}
\usepackage{subfig}
\usepackage{color}
\usepackage{algorithmic}
\usepackage{graphicx}
\usepackage{textcomp}
\usepackage{xcolor}
\usepackage{xspace}
\usepackage{soul}
\usepackage{marginnote}
\usepackage{longtable}
\usepackage[framemethod=default]{mdframed}
\usepackage{pdflscape}
\usepackage{pifont}
\usepackage{adjustbox}
\usepackage[hidelinks]{hyperref}
\usepackage[T1]{fontenc}
\usepackage{subfig}
\usepackage{xcolor}

\newcommand{\cmark}{\ding{51}}%
\newcommand{\xmark}{\ding{55}}%

\newcommand{\ie}{\emph{i.e.,}\xspace}
\newcommand{\eg}{\emph{e.g.,}\xspace}
\newcommand{\etc}{etc.\xspace}
\newcommand{\etal}{\emph{et~al.}\xspace}

\newcommand{\CHANGED}[1]{#1}

\definecolor{bleudefrance}{rgb}{0.19, 0.55, 0.91}

\newcommand{\replicationPackage}
{\url{https://doi.org/10.6084/m9.figshare.24603237}}

\journal{Journal of Systems and Software}

\begin{document}

\begin{frontmatter}



\title{Incivility Detection in Open Source Code Review and Issue Discussions}

\author[1]{Isabella Ferreira\corref{cor1}}
\ead{isabella.ferreira@polymtl.ca}
\cortext[cor1]{Corresponding author}

\author[2]{Ahlaam Rafiq}
\ead{arafiq@iitg.ac.in}

\author[1]{Jinghui Cheng}
\ead{jinghui.cheng@polymtl.ca}

\address[1]{Department of Computer and Software Engineering, Polytechnique Montréal, Montréal, Quebec, Canada}
            
\address[2]{Department of Physics, Indian Institute of Technology, Guwahati, Assam, India}


\begin{abstract}
Given the democratic nature of open source development, code review and issue discussions may be uncivil. Incivility, defined as features of discussion that convey an unnecessarily disrespectful tone, can have negative consequences to open source communities. To prevent or minimize these negative consequences, open source platforms have included mechanisms for removing uncivil language from the discussions. However, such approaches require manual inspection, which can be overwhelming given the large number of discussions. To help open source communities deal with this problem, in this paper, we aim to compare six classical machine learning models with BERT to detect incivility in open source code review and issue discussions. Furthermore, we assess if adding contextual information in \CHANGED{the previous email/comment} improves the models’ performance and how well the models perform in a cross-platform setting. We found that BERT performs better than classical machine learning models, with a best F1-score of 0.95. Furthermore, classical machine learning models tend to underperform to detect tone-bearing and civil discussions. Our results show that adding \CHANGED{the previous email/comment} to BERT did not improve its performance and that none of the analyzed classifiers had an outstanding performance in a cross-platform setting. Finally, we provide insights into the tones that the classifiers misclassify and lessons learned for using automated techniques in incivility detection.
\end{abstract}




\end{frontmatter}


\section{Introduction}

Open source software (OSS) development provides abundant opportunities for public discussions, which happen within the context of issue tracking, bug report, code review, and user feedback, to just name a few. These opportunities characterize the democratic essence of open source development by allowing anyone who has the relevant knowledge to contribute to the development process and shape the project one way or another. However, as in all types of public discussions, conversations that happen in open source development can become uncivil. Take, for example, this code review comment of a patch submitted to the Linux kernel: ``\textit{What the F*CK, guys? This piece-of-shit commit is marked for stable, but you clearly never even test-compiled it, did you?}'' Although the comment expressed opinions on a technical issue, the commenter used an unnecessarily disrespectful tone.

We define \textit{incivility}, in the context of software engineering (SE) in general and OSS development in specific, as \textit{features of discussion that convey an unnecessarily disrespectful tone toward the discussion forum, its participants, or its topics}~\cite{ferreira2021shut}. This concept is related to other constructs used to describe unhealthy discussions in the context of SE, such as \textit{hate speech}, \textit{offensive language}, \textit{toxicity}, and \textit{pushback}. Although these different concepts might share certain characteristics with incivility, in our definition, the concept of incivility covers a broader spectrum~\cite{ferreira2021shut}. That is, the construct of toxicity often emphasizes the \textit{impact} on the target of incivility; \eg \textit{``make someone leave a discussion''}~\cite{miller2022did} and \textit{``impact the health of FOSS/peer production communities''}~\cite{carillo2016dose}. Yet, the concept of incivility is not confined to this aspect. While hate speech and offensive language usually focus on a \textit{specific emotion/tone} (such as entitlement~\cite{ramanstress}, insults~\cite{cheriyan2021towards}, or racist terms~\cite{davidson2017automated}), incivility does not~\cite{ferreira2021shut}. Finally, while both incivility and pushback are unnecessary behaviors, pushback focuses on a specific action (\ie ``\textit{a reviewer blocking a change request}''~\cite{egelman2020predicting}), while incivility is broader. Based on the aforementioned differences, \CHANGED{we argue that compared to the other constructs, the concept of incivility is not confined to a particular impact or specific emotions/tones and thus is more general.}

Although incivility can be rare, uncivil expressions often have important impacts on the communication and the discussion participants, resulting in escalated incivility, discontinued conversation, or disengaged contributors~\cite{ferreira2021shut}. As such, many major SE platforms such as Stack Overflow and GitHub have incorporated mechanisms for labeling and removing offensive and toxic languages~\cite{heat_detector, holman_2014}. Many of these approaches involve manual inspection, which requires considerable human efforts given the large amount of content generated daily in those platforms. Additionally, they often target one construct and ignores the broader spectrum of uncivil discussion features. Hence, automated techniques for detecting uncivil communication in software engineering platforms would help open source communities to proactively manage uncivil interactions.

Several tools have been developed by researchers to detect toxicity~\cite{ramanstress, sarker2020benchmark, sarker2023automated}, offensive language~\cite{cheriyan2021towards}, and pushback~\cite{egelman2020predicting}. However, these tools do not perform well when it comes to new samples or SE discussions~\cite{miller2022did, ramanstress, sarker2020benchmark, qiu2022detecting}. Additionally, the concept of incivility may cover a wider spectrum than those previously explored constructs~\cite{ferreira2021shut}. Based on these results, we were motivated to build an incivility-specific classifier.

Developing such a classifier, however, involves major challenges. First, although with great impacts, uncivil exchanges in open source communities can be infrequent. Previous studies identified that only 7.25\% of code review comments~\cite{ferreira2021shut} and 8.82\% of issue comments~\cite{ferreira2022how} demonstrate incivility. The lack of uncivil cases poses challenges in creating datasets for training and evaluating the automated techniques. Second, incivility can be manifested in various ways. For example, previous work has identified many characteristics of discussion that can be seen as uncivil. Among them, there are straightforward features such as name calling and vulgarity. But at the same time, incivility can be manifested through discussion characteristics such as irony, mocking, and threat that are difficult to detect automatically~\cite{ferreira2021shut}. As a result, many existing software engineering sentiment analysis tools do not perform well when detecting incivility~\cite{ferreira2021shut}. Finally, incivility can be ``very much in the eye of the beholder''~\cite{coe2014online}. Thus, the discussion context can have strong indications on whether a comment is uncivil. So analyzing the text in isolation may lead to inaccurate results.

We aimed to address these challenges in our research by exploring machine learning and deep learning techniques trained with two manually labeled datasets~\cite{ferreira2021shut, ferreira2022how}. Particularly, we investigated data augmentation and class balancing techniques to facilitate the detection of infrequent uncivil comments. The datasets we used are also created by considering the complexity of civil and uncivil discussion features, such as humility, confusion, sadness, irony, and mocking. Additionally, by exploring how \CHANGED{the previous discussion} can affect the model performance, we aimed to incorporate the beholder’s perspective, addressing the problem of classifying the text in isolation. More especially, we pose the following three research questions.\\

\noindent
\textbf{RQ1. How well can machine learning models (including BERT-based model and classical machine learning models) detect incivility?}

To the best of our knowledge, none of the previous research has built classifiers to detect incivility in open source code review or issue discussions. Hence, it is unknown if incivility can be automatically detected with a good performance in such discussions. Previous studies have identified that BERT-based deep learning models~\cite{devlin2018bert} outperformed classical machine learning models when detecting related concepts such as toxicity~\cite{sarker2023automated} and offensive language~\cite{cheriyan2021towards}. Based on these results, we hypothesize that BERT can detect incivility in open source code review and issue discussions with satisfying performance and can outperform the classical machine learning models.\\

\CHANGED{\noindent
\textbf{RQ2. To what extent does adding the previous email/comment help to detect incivility in code review and issue discussions?}

Ferreira \etal~\cite{ferreira2021shut} have shown that the existing discussion and the context are important factors that should be considered when detecting incivility in code review discussions. However, Murgia \etal~\cite{murgia2014developers} found that adding previous discussions did not help human raters reach an agreement on assessing emotion expressed in comments on issue tracking systems. Considering these conflicting conclusions in previous work, we aim to assess if adding the previous discussion helps to improve the performance of automated incivility detection techniques.\\}

\noindent
\textbf{RQ3. How well do the incivility detection techniques work in a cross-platform setting?}

Building a manually annotated gold standard for incivility detection on a particular platform is a time-consuming task. Currently, only two datasets are available on the literature~\cite{ferreira2021shut, ferreira2022how}, focusing on code review discussions and issue discussions, respectively. However, discussions that happened on different platforms could have characteristics that indicate incivility in different ways. Hence, in RQ3, we aim to assess if it is feasible to use BERT and classical machine learning models to detect incivility in a cross-platform setting. This information will help us assess the performance of incivility detection on a new dataset when a gold standard is not available.\\

\CHANGED{\noindent
\textbf{RQ4: What types of discussion features tend to be misclassified by the incivility detection techniques?}

Incivility is a broad and complex construct that can be manifested by different discussion features such as \textit{irony}, \textit{mocking}, and \textit{bitter frustration}~\cite{ferreira2021shut,ferreira2022how}. The performance of the automated detectors may vary when detecting incivility demonstrating different features. Thus in RQ4, we investigate the performance of incivility detection techniques on different discussion features. The goal of this analysis is to identify error patterns and further inform the creation of practical tools that address limitations of automated techniques.\\}

Similar to our previous work of manual incivility labeling~\cite{ferreira2021shut,ferreira2022how}, our incivility detection approach follows two steps. First, SE discussions are classified into \textit{tone-bearing} and \textit{non-tone-bearing} groups; the former includes discussions that demonstrate at least one tone-bearing discussion feature (such as sadness, appreciation, or frustration) while the latter includes other discussions that are often purely technical. Then, we classify the tone-bearing discussions into \textit{civil} or \textit{uncivil} depending on the specific discussion feature that they demonstrate~\cite{ferreira2021shut}. 

Our results show that BERT outperforms classical machine learning techniques, having an outstanding performance when detecting incivility in both code review and issue discussions ($F1>0.9$). Additionally, we found that classical machine learning models tend to underperform to classify non-tone-bearing comments as well as civil sentences in both datasets. Our findings also indicated that adding \CHANGED{the previous email/comment} did not improve the incivility classification. Furthermore, we found that although none of the classifiers have had an outstanding performance to detect incivility in a cross-platform setting, all classifiers were able to perform well when detecting the non-tone-bearing and uncivil classes. \CHANGED{Our analysis of misclassified cases also shed light on the complexity of incivility detection when considering the various features of discussions.} Together, our results allow us to summarize lessons learned for developing incivility detection techniques. The dataset, source code, and evaluation results are available at our replication package\footnote{\replicationPackage}.

\section{Background \& Related Work}

In this section, we discuss the background information and related work on (i) machine learning for text classification, (ii) automated detection of unhealthy discussions in online communication platforms, and (iii) automated detection of unhealthy discussions in software engineering. 

\subsection{Machine Learning for Text Classification}

Text classification is a classical natural language processing (NLP) problem that aims at assigning labels to textual documents, such as sentences or paragraphs~\cite{minaee2021deep}. Currently, there are two kinds of machine learning approaches for automatic text classification, namely \textit{classical machine learning-based models} and \textit{neural network-based approaches}.

Common \textbf{classical machine learning-based models} include \textit{classification and regression tree (CART)}, \textit{k-nearest-neighbors (KNN)}, \textit{logistic regression}, \textit{naive Bayes}, \textit{random forest}, and \textit{support vector machine (SVM)}, among others. They were applied in various general text classification tasks~\cite{shah2020comparative, pranckevivcius2017comparison, dai2007transferring, lewis2000introduction}, as well as for software engineering tasks in specific~\cite{rahman2019source, arya2019analysis, chouchen2021predicting, uchoa2021predicting}. To use these models, features need to be first defined and extracted from textual documents, then fed into the classifier for prediction. Popular features for textual data include \textit{bag of words (BoW)} and \textit{term frequency–inverse document frequency (tf-idf)}. Although widely used, classical machine learning classifiers have a major limitation. That is, choosing the proper features for each domain requires extensive domain knowledge; thus it is hard to define cross-domain or cross-task features~\cite{minaee2021deep}. In this work, we assess if it is feasible to use the six aforementioned classical machine learning-based models to detect incivility in code review and issue discussions.

To solve the aforementioned challenges, \textbf{neural network-based approaches} have been widely explored in the literature to address text classification tasks~\cite{devlin2018bert, bengio2000neural, mikolov2013distributed, ilic2018deep, vaswani2017attention, radford2018improving, brown2020language}. In 2018, Devlin \etal~\cite{devlin2018bert} proposed \textit{BERT (Bidirectional Encoder Representations from Transformers)}, which is currently the state of the art embedding model~\cite{minaee2021deep}. The BERT base model consists of 110M parameters and has been trained on BookCorpus~\cite{bookcorpus} and English Wikipedia~\cite{wikipedia}, which include a total of 3.3 billion words. 
BERT is trained with two objectives: masked language modeling (MLM) and next sentence prediction (NSP). MLM allows the model to learn a bidirectional representation of the sentence by randomly masking 15\% of the words in the input and then training the model to predict the masked words. For NSP, the model concatenates two masked sentences as inputs during the pretraining phase and then predicts if the two sentences are continuous in the text or not. 

Many variants have been made to BERT since it was proposed~\cite{liu2019roberta, lan2019albert, sanh2019distilbert, joshi2020spanbert, feng2020codebert}. RoBERTa~\cite{liu2019roberta}, for example, is a more robust implementation of BERT, trained with a much larger amount of data. ALBERT~\cite{lan2019albert} optimizes BERT by lowering its memory consumption and increasing its training speed. DistillBERT~\cite{sanh2019distilbert} uses knowledge distillation, \ie a compression technique in which a compact model is trained to reproduce the behavior of a larger model, to reduce the size of the BERT model. SpanBERT~\cite{joshi2020spanbert} is a pre-trained method to better represent and predict spans of text. CodeBERT~\cite{feng2020codebert} is a pre-trained language model for both programming languages and natural languages. In many NLP~\cite{gonzalez2020comparing} and software engineering problems~\cite{cheriyan2021towards, biswas2020achieving, batra2021bert, wu2021bert}, BERT has demonstrated to have better performance than classical machine learning models. Thus, we investigate BERT's ability to detect incivility in code review and issue discussions. To simplify this initial exploration, we used the original BERT model instead of its variants.

\subsection{Automated Detection of Incivility in Online Communication Platforms}

By using either classical machine learning-based models or neural network-based approaches, many authors have tried to automatically detect incivility on online platforms, such as in news discussions~\cite{daxenberger2018automatically, sadeque2019incivility} and Twitter~\cite{maity2018opinion}. Daxenberge \etal~\cite{daxenberger2018automatically}, for example, sought to understand incivility (defined as ``\textit{expressions of disagreement by denying and disrespecting opposing views}'') on user comments of nine German media outlets on Facebook. By using a logistic regression classifier, they found that incivility can be identified with an overall F1-score of 0.46. To assess how well machine learning models are able to detect incivility (defined as ``\textit{features of discussion that convey an unnecessarily disrespectful tone towards the discussion forum, its participants, or its topics}'') in a cross-platform setting, Sadeque \etal~\cite{sadeque2019incivility} trained different machine learning models on an annotated newspaper dataset and tested them on Russian troll tweets. As a result, Recurrent Neural Network (RNN) with Gated Recurrent Units (GRU) outperformed the other analyzed models with an F1-score of 0.51 for name calling and 0.48 for vulgarity. On Twitter, incivility (defined as ``\textit{the act of sending or posting mean text messages intended to mentally hurt, embarrass or humiliate another person using computers, cell phones, and other electronic devices}'') detection with character-level bidirectional long short-term memory (bi-LSTM) and character-level convolutional neural networks (CNNs) with a rectified linear unit (ReLU) outperformed the best baseline model with a F1-score of 0.82~\cite{maity2018opinion}.

In our literature review, we were not able to find previous research investigating automated detection of incivility in software engineering settings, although some recent work focused on detecting unhealthy discussions that we review in the next section. In this paper, we thus address this gap by leveraging the concepts and the datasets established in our previous work about incivility in code review~\cite{ferreira2021shut} and issue discussions~\cite{ferreira2022how}. 

\subsection{Automated Detection of Unhealthy Discussions in Software Engineering}

Unhealthy interactions are often characterized in software engineering (SE) discussions as  \textit{toxicity}~\cite{miller2022did, carillo2016dose, sarker2023automated, carillo2016towards, qiu2022detecting}, \textit{offensive language}~\cite{cheriyan2021towards, sarker2020benchmark}, \textit{heated discussions}~\cite{ferreira2022how, rahman2019source}, \textit{hate speech}~\cite{ramanstress}, and \textit{pushback}~\cite{egelman2020predicting, qiu2022detecting}. Table~\ref{table:automated_detection_se} presents the studies proposing models to detect different kinds of unhealthy interactions. We compare our study with the literature with respect to the model implemented, the used dataset, and the techniques to improve the models' performance, such as cross-validation (\ie iteratively training and testing a model using different portions of the data), data augmentation (\ie artificially increasing data points in the training set by generating new data points from existing ones), class balancing (\ie adjusting the class sizes with unbalanced samples), and hyperparameter optimization (\ie choosing the hyperparameter combination for optimum performance).

\begin{table*}[!]
\centering
\caption{Methods available in the literature to automatically detect unhealthy discussions in the software engineering domain.}
\label{table:automated_detection_se}
\begin{adjustbox}{angle=90}
\scalebox{0.65}{
\begin{tabular}{l|l|l|l|l|l}
\cellcolor[HTML]{D9D9D9}\textbf{Authors} &
\cellcolor[HTML]{D9D9D9}  \textbf{Goal} &
\cellcolor[HTML]{D9D9D9} \textbf{Model} &
\cellcolor[HTML]{D9D9D9}\textbf{ Dependent variables} &
\cellcolor[HTML]{D9D9D9}  \textbf{Dataset} &
\cellcolor[HTML]{D9D9D9}  \textbf{Techniques} 
\\ \toprule
  
  
Schneider \etal~\cite{schneider2016differentiating}  
& 
  \begin{tabular}[c]{@{}l@{}}Identify the \textbf{discourse patterns} of the \\ leaders of the LKML.\end{tabular} &
  Naive Bayes &
  \begin{tabular}[c]{@{}l@{}}Email sent by Linus Torvalds\\ Email sent by Greg Kroah-Hartman\end{tabular} &
 Code reviews 
  &  
  	 \begin{tabular}[c]{@{}l@{}}
	Cross-validation: \cmark   \\
	Data augmentation: \xmark \\
	Class balancing: \xmark \\
	Hyperparameter optimization:  \xmark 
	\end{tabular}
	
\\ \midrule
  
  
Gachechiladze \etal~\cite{gachechiladze2017anger} 
&
  \begin{tabular}[c]{@{}l@{}}Detect \textbf{anger} towards self, others, and \\ objects.\end{tabular} &
  Weka implementation: SVM, J48, Naive Bayes &
  \begin{tabular}[c]{@{}l@{}}Anger towards self\\ Anger towards others\\ Anger towards objects\end{tabular} &
  Jira issues &  
  	 \begin{tabular}[c]{@{}l@{}}
	Cross-validation: \cmark   \\
	Data augmentation: \xmark \\
	Class balancing: \xmark \\
	Hyperparameter optimization:  \cmark 
	\end{tabular}

\\ \midrule
  
    

  
    
Egelman \etal~\cite{egelman2020predicting} 
&
  \begin{tabular}[c]{@{}l@{}}Detect the feelings \textbf{pushback} in code \\ reviews, \ie the perception of unnecessary \\ interpersonal conflicts in code review while \\ a reviewer is blocking a change request.\end{tabular} &
  Logit Regression Model &
  \begin{tabular}[c]{@{}l@{}} Interpersonal conflict\\ Feeling that acceptance was withheld for too long\\ Reviewer asked for excessive changes\\ Feeling negative about future code reviews\\ Frustration\end{tabular} &
  Code reviews  & 
    	 \begin{tabular}[c]{@{}l@{}}
	Cross-validation: \xmark   \\
	Data augmentation: \xmark \\
	Class balancing: \xmark \\
	Hyperparameter optimization:  \xmark 
	\end{tabular}
  \\ \midrule
  
  
Raman \etal~\cite{ramanstress} 
&
  \begin{tabular}[c]{@{}l@{}}Detect \textbf{toxic language}, \ie hate speech \\ and microaggressions.\end{tabular} &
  SVM &
  \begin{tabular}[c]{@{}l@{}}Toxic\\ Non-toxic\end{tabular} &
  GitHub issues &
    	 \begin{tabular}[c]{@{}l@{}}
	Cross-validation: \cmark   \\
	Data augmentation: \xmark \\
	Class balancing: \xmark \\
	Hyperparameter optimization:  \cmark 
	\end{tabular}
\\ \midrule
  
  
Sarker \etal \cite{sarker2020benchmark} 
&
  Evaluate different tools to detect \textbf{toxicity}. &
  \begin{tabular}[c]{@{}l@{}}Perspective API, STRUDEL Toxicity Detector, \\ Deep Pyramid Convolutional Neural  Networks, \\ BERT with fast.ai, \\Hate Speech Detection\end{tabular} &
  \begin{tabular}[c]{@{}l@{}}Toxic\\ Non-toxic\end{tabular} &
  \begin{tabular}[c]{@{}l@{}} Code reviews, \\ Gitter messages \end{tabular} &
    	 \begin{tabular}[c]{@{}l@{}}
	Cross-validation: \xmark   \\
	Data augmentation: \xmark \\
	Class balancing: \xmark \\
	Hyperparameter optimization:  \xmark 
	\end{tabular}
\\ \midrule

Cheriyan \etal~\cite{cheriyan2021towards} 
&
  \begin{tabular}[c]{@{}l@{}} Detect and classify \textbf{offensive language}, \\ \ie communication that contains gutter \\ language, swearing, racist, or offensive \\ content. \end{tabular} &
  Random Forest, SVM, BERT &
\begin{tabular}[c]{@{}l@{}} Offensive \\ Non-offensive \end{tabular} 
  &
  \begin{tabular}[c]{@{}l@{}}  GiHub, \\ Gitter, \\ Slack, \\ Stack Overflow\end{tabular}  &
  
      \begin{tabular}[c]{@{}l@{}}
	Cross-validation: \xmark   \\
	Data augmentation: \cmark \\
	Class balancing: \xmark \\
	Hyperparameter optimization:  \xmark 
	\end{tabular}
  
  \\ \midrule
  
  
 

  
    
     Sarker \etal~\cite{sarker2023automated}
&
  \begin{tabular}[c]{@{}l@{}}Detect \textbf{toxicity}, \ie SE conversations\\  that include offensive, name calling, \\ insults, threats, personal attacks, \\ flirtations, reference to sexual activities, \\ and swearing or cursing.\end{tabular} &
  
  \begin{tabular}[c]{@{}l@{}} Decision tree, Logistic Regression, SVM, \\ Random Forest, Gradient-Boosted Decision Trees, \\ LST, BiLSTM, GRU, DPCNN, BERT \end{tabular} &
  
  \begin{tabular}[c]{@{}l@{}}Toxic\\ Non-toxic \end{tabular} &
  
  \begin{tabular}[c]{@{}l@{}}Code reviews \\ Gitter \end{tabular} &
  
  
     \begin{tabular}[c]{@{}l@{}}
	Cross-validation: \cmark   \\
	Data augmentation: \xmark \\
	Class balancing: \xmark \\
	Hyperparameter optimization:  \cmark 
	\end{tabular}
    
    \\ \midrule
       
\rowcolor[HTML]{D9D9D9} Our work
&
  \begin{tabular}[c]{@{}l@{}}Detect \textbf{incivility}, \ie features of discussion \\ that convey an unnecessarily disrespectful \\ tone toward the discussion forum, its \\ participants, or its topics in code review and \\ issue discussions.\end{tabular} &
  
  \begin{tabular}[c]{@{}l@{}}CART, KNN, Logistic Regression,\\  Naive Bayes, Random Forest,  SVM, BERT\end{tabular} &
  
  \begin{tabular}[c]{@{}l@{}}Tone-bearing\\ Non-tone-bearing\\ Civil\\ Uncivil\end{tabular} &
  
  \begin{tabular}[c]{@{}l@{}}Code reviews \\ GitHub issues \end{tabular} &
  
  
     \begin{tabular}[c]{@{}l@{}}
	Cross-validation: \cmark   \\
	Data augmentation: \cmark \\
	Class balancing: \cmark \\
	Hyperparameter optimization:  \cmark 
	\end{tabular}

  \\ \bottomrule
\end{tabular}%
}
\end{adjustbox}
\end{table*}

Previous studies identified that open source contributors might have different, sometimes negative, communication styles. For example, a Naive Bayes classifier identified that the leaders of the Linux Kernel Mailing List (LKML) have different communication styles (F1-score = 0.96)~\cite{schneider2016differentiating}; some used more impolite, rude, aggressive, or offensive words. Offensive language (defined as ``\textit{communication that contains gutter language, swearing, racist, or offensive content''}) can also be identified in other platforms such as GitHub, Gitter, Slack, and Stack Overflow, with more than 97\% of accuracy using BERT~\cite{cheriyan2021towards}.

In addition to having different communication styles, contributors might also demonstrate negative emotions when expressing themselves in open source discussions. Anger, for example, can be accurately identified in Jira discussions with SVM (F1-score = 0.81), J48 decision tree (F1-score = 0.77), and Naive Bayes (F1-score = 0.72)~\cite{gachechiladze2017anger}. Another negative emotion that might emerge in code review discussions, more specifically, is the feeling of pushback, which is characterized by interpersonal conflicts, impatience, disappointment, and frustration~\cite{egelman2020predicting}. In Google's code review discussions, a logistic regression model found that code review authors are between 3.0 and 4.1 times more likely to experience the feeling of pushback for at least once and between 7.0 and 13.7 times more likely to experience it multiple times when compared to code reviews that were not flagged with a potential feeling of pushback.

Finally, toxic language, \ie hate speech and microaggressions, can be identified with automated methods. For example, using the SVM model Raman \etal detected toxicity in GitHub issues with a precision of 0.75, but a low recall of 0.35~\cite{ramanstress}. Sarker et al.~\cite{sarker2020benchmark} also tested the SVM classifier on other 100k randomly sampled GitHub issues; they found that the precision decreased to 0.50, demonstrating that the model might be overfitting to the training set. Similarly, toxicity can be identified in Gerrit code review and Gitter discussions, with the STRUDEL toxicity detector having an F1-score of 0.49 and 0.73, respectively~\cite{sarker2020benchmark}. Interestingly, Sarker \etal~\cite{sarker2020benchmark} found that toxicity detectors tend to perform worse on more formal SE discussions, such as code reviews, than on informal conversations such as Gitter messages. Finally, a recent study~\cite{sarker2023automated} proposes ToxiCR, ten supervised machine learning algorithms with a combination of text vectorization and processing steps on 19,651 code review comments.
ToxiCR has an F1-score of 0.89. 

Our work differs from the previous works in several ways. First, our study proposes an incivility-specific classifier for SE discussions. This study builds upon our previous work on characterizing incivility in code review discussions of rejected patches from the LKML and GitHub issue discussions locked as too heated. We chose to compare six classical machine learning models (\textit{Classification and Regression Tree (CART)}, \textit{k-Nearest Neighbors (KNN)}, \textit{Logistic Regression}, \textit{Naive Bayes}, \textit{Random Forest}, and \textit{SVM}) with \textit{BERT}. Additionally, we use four strategies to augment our data (\ie \textit{synonym replacement}, \textit{random insertion}, \textit{random swap}, and \textit{random deletion}) and compare three class balancing techniques, \ie \textit{random undersampling}, \textit{random oversampling}, and \textit{SMOTE}. We also perform hyperparameter optimization with \textit{Grid Search} on the classical machine learning models' hyperparameters and \textit{Bayesian Optimization} on BERT's hyperparameters to improve the models' performance. Finally, the performance of our models is evaluated in a \textit{5-fold cross-validation}. On top of evaluating the performance of the machine learning models in detecting incivility, we also analyze the impact of \CHANGED{the previous discussion} and the feasibility of detecting incivility in a cross-platform setting. Finally, we assess what are the different kinds of uncivil discussion features that the models are able to (mis)classify.

\section{Study Design}

\begin{figure*}[ht]
\centering
\includegraphics[clip, width=0.95\linewidth]{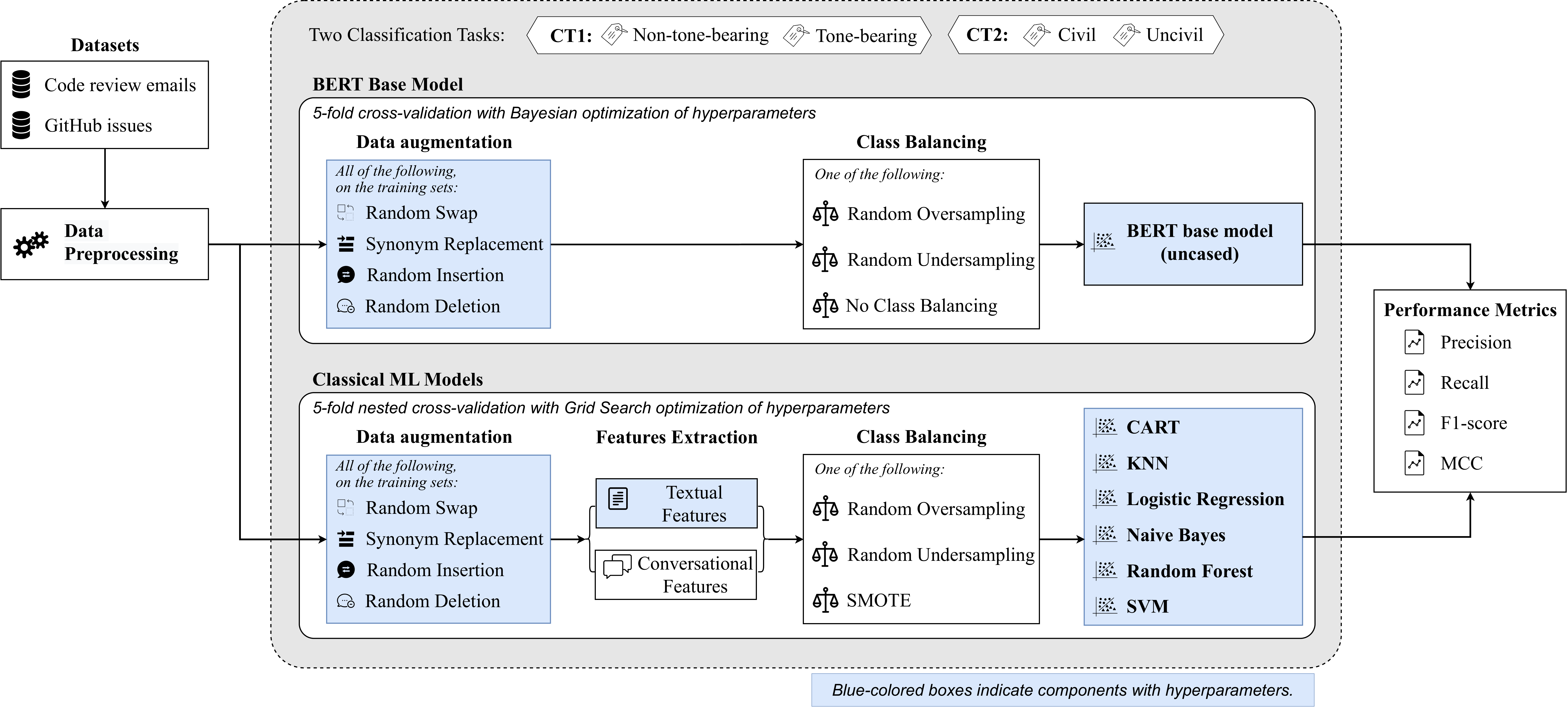}
\caption{Key components and main pipeline of incivility classifiers}
\label{fig:methodology_supervised}
\end{figure*}

\CHANGED{Figure~\ref{fig:methodology_supervised} depicts the key components in the pipeline of the classifiers explored in this study. After preprocessing the datasets (Sections~\ref{section:dataset} and~\ref{sec:data_preprocessing}) and extracting the features for the classical machine learning models (Section~\ref{section:feature_extraction}), we stratify our dataset into train, test, and validation sets. Then, we augment our training set (Section~\ref{section:eda}) and balance our classes (Section~\ref{section:class_balancing_techniques}). We investigate six classical machine learning models (Section~\ref{section:classifiers}) and one deep learning model (Section~\ref{section:bert}). During training, to obtain the optimal models, we perform hyperparameter tuning to find the best set of hyperparameters. We then test the trained classifiers on the test set and assess the performance of each classifier according to four performance metrics (Section~\ref{section:performance_metrics}). Finally, we perform experiments and additional analysis to answer the four RQs (Section~\ref{sec:methodology:rqs}).}

\subsection{Datasets}
\label{section:dataset}

The general goal of this study is to assess the extent to which incivility can be detected in code reviews and issue discussions. To the best of our knowledge, only two incivility datasets are available in the literature, \ie a \textbf{code review dataset} comprising code review emails of rejected patches that were sent to Linux Kernel Mailing List (LKML)~\cite{ferreira2021shut} and an \textbf{issues dataset} comprising GitHub issues locked as too heated~\cite{ferreira2022how}. We used both datasets to train our classifiers.

For each dataset, the natural language emails (in the case of the code review dataset) and comments (in the case of the GitHub issues dataset) were first labeled as \textit{non-tone-bearing} or \textit{tone-bearing}. Following the definition used by the datasets, the \textbf{non-tone-bearing} class (formerly called \textit{technical} by Ferreira \etal~\cite{ferreira2021shut}) comprises emails and comments in which none of their sentences convey a mood or style of expression~\cite{ferreira2021shut}. On the contrary, \textbf{tone-bearing} (previously called \textit{non-technical} by Ferreira \etal~\cite{ferreira2021shut}) code review emails or issue comments are those in which at least one sentence expresses a tone-bearing discussion feature (TBDF). The datasets use the concept of TBDF to indicate ``\textit{conversational characteristics demonstrated in a written sentence that convey a mood or style of expression'}'~\cite{ferreira2021shut}. These TBDFs are further divided into four categories: (1) civil positive (e.g., \textit{humility} and \textit{excitement}), (2) civil neutral (e.g., \textit{apologies} and \textit{friendly joke}), (3) civil negative (e.g., \textit{sadness} and \textit{oppression}), and (4) uncivil (e.g., \textit{bitter frustration}, \textit{impatience}, \textit{mocking}, or \textit{vulgarity})~\cite{ferreira2021shut, ferreira2022how}. In total, there are 1,365 non-tone-bearing emails and 168 tone-bearing emails in the code review dataset; there are 4,793 non-tone-bearing comments and 718 tone-bearing comments in the issues dataset.

Next, sentences in tone-bearing emails and comments were then further categorized as \textit{civil} or \textit{uncivil}. The \textbf{uncivil} class contains sentences that demonstrate at least one uncivil TBDF as defined in the dataset~\cite{ferreira2021shut, ferreira2022how}. Conversely, \textbf{civil} sentences are those that only contain civil TBDFs (either civil positive, civil neutral, or civil negative, as described above). There are 117 civil sentences and 276 uncivil sentences in the tone-bearing emails of the code review dataset and there are 353 civil sentences and 896 uncivil sentences in the tone-bearing comments of the issues dataset.

\CHANGED{Our classification tasks are thus two-layered to mimic the human classification process in our previous work~\cite{ferreira2021shut}}: first, we aim to \textbf{classify code review emails and issue comments into tone-bearing or non-tone-bearing (CT1)}; then, for tone-bearing contents, we aim to \textbf{classify sentences into civil or uncivil (CT2)}. The goal to separate these two classification tasks is because in a concrete scenario in which open source contributors would use our classifiers to assess whether their comments are uncivil or not, first, we would detect if it is a tone-bearing text or not. If it is tone-bearing, then we would detect (in)civility.

\subsection{Data Preprocessing} 
\label{sec:data_preprocessing}

We followed a series of steps to reduce noise on the datasets described above. First, we considered sentences coded with both civil and uncivil TBDFs as uncivil data points. This is because the appearance of any uncivil expression can already signal an unhealthy discussion, regardless of the appearance of civil tones~\cite{ferreira2021shut}; for example, ``\textit{I don't think `not fixing it because it's not fixed yet' is a good reason to keep things the way they are. But maybe that's just me.''} should be considered as an uncivil sentence for demonstrating \textit{impatience} although a civil tone of \textit{humility} is present. We then performed the following preprocessing steps:

\begin{enumerate}[leftmargin=*,nosep]
	\item We manually remove the source code (including variable names, function names, stack traces, \textit{etc.}), words other than English, emojis, and GitHub username mentions (such as @username) from the text;
	
	\item We automatically remove the header of code review emails, including the first line that follows the regex pattern ``\texttt{{On (.*?) wrote:}}'';
	
	\item We automatically remove greetings such as ``Hi [person\_name]'' and statements such as ``Reviewed by [person\_name]'' or ``Tested by [person\_name]'';
	
	\item We automatically remove any signature statement that is in the following list of words: ``warm regards'', ``kind regards'', ``regards'', ``cheers'', ``many thanks'', ``thanks'', ``sincerely'', ``best'', ``thank you'', ``talk soon'', ``cordially'', ``yours truly'', ``all the best'', ``best regards'', ``best wishes'', ``looking forward to hearing from you'', ``sincerely yours'', ``thanks again'', ``with appreciation'', ``with gratitude'', and ``yours sincerely'';                                                
    \item We automatically remove all reply quotes, usually represented by ``\textless'';
    
    \item We automatically remove stop words and punctuation; we perform stemming on each remaining word. This step is only performed for the classical machine learning classifiers.
	
 \end{enumerate}

\subsection{Feature Extraction for Classical ML Classifiers}
\label{section:feature_extraction}

This step consists in extracting features to detect incivility in code review emails/issues comments and sentences using supervised techniques. We created two sets of features for both classification tasks, namely \textit{textual features} and \textit{conversational features}, which were inspired by the work of Arya \etal~\cite{arya2019analysis} and adapted to our context.

\begin{table*}[ht!]
\centering
\caption{Conversational features of code review and issue discussions}
\label{table:features}
\resizebox{\textwidth}{!}{%
\begin{tabular}{l|l|l|l|l}
\multicolumn{1}{c|}{\cellcolor[HTML]{D9D9D9}\textbf{Feature type}} &
  \multicolumn{1}{c|}{\cellcolor[HTML]{D9D9D9}\textbf{Classification task}} &
  \multicolumn{1}{c|}{\cellcolor[HTML]{D9D9D9}\textbf{Feature name}} &
  \multicolumn{1}{c|}{\cellcolor[HTML]{D9D9D9}\textbf{Description}} &
  \multicolumn{1}{c}{\cellcolor[HTML]{D9D9D9}\textbf{Values}} \\ \toprule
Participant &
  CT1, CT2 &
  AUTHOR\_ROLE &
  \begin{tabular}[c]{@{}l@{}}Email author's role in the Linux kernel. We first group identities that have the same names or the same email addresses.
  The \\ author is considered a maintainer if one of those identities appears in the MAINTAINERS file~\cite{maintainers_file}, and a developer otherwise.\end{tabular} &
  \{Maintainer, Developer\} \\
\multirow{5}{*}{Length} &
  CT1, CT2 &
  FIRST\_AUTHOR &
  Flag if the email/comment author also sent the first email/comment of thread (\ie original patch/issue description). &
  \{True, False\} \\ \midrule
 &
  CT1 &
  CHAR\_TEXT &
  Number of characters in the email/comment. &
$\mathbb{R}_{ \geq 0} = \{{x \in \mathbb{R}} | x_{\geq 0}\} $ \\
 &
  CT1 &
  LEN\_TEXT &
  Number of words in the email/comment divided by that of the longest email/comment in the thread. &
  (0,1{]} \\
 &
  CT2 &
  CHAR\_SENT &
  Number of characters in the sentence. &
  (0,1{]} \\
 &
 CT2 &
  LEN\_SENT\_T &
  Number of words in sentence divided by that of longest sentence in the thread. &
  (0,1{]} \\
  &
  CT2 &
  LEN\_SENT\_C &
  Number of words in sentence divided by that of longest sentence in the email/comment. &
  (0,1{]} \\ \midrule
\multirow{4}{*}{Structural} &
  CT1 &
  POS\_TEXT\_T &
  Position of email/comment in the thread divided by the number of emails/comments in thread. &
  (0,1{]} \\
 &
  CT2 &
  POS\_SENT\_E &
  \begin{tabular}[c]{@{}l@{}}Position of sentence in email/comment divided by the number of sentences in email/comment. Sentences are identified based \\ on the following regular expression: $(?<=[.!?])$.\end{tabular}
  &
  (0,1{]} \\
 &
  CT2 &
  POS\_SENT\_T &
  \begin{tabular}[c]{@{}l@{}}Position of sentence in thread divided by the number of sentences in thread. Sentences are identified based on the following \\ regular expression: $(?<=[.!?])$.\end{tabular} &
  (0,1{]} \\
 &
  CT1, CT2 &
  LAST\_COMMENT &
  Flag if it is the last email/comment or not. &
  \{True, False\} \\ \midrule
\multirow{4}{*}{Temporal} &
  CT1, CT2 &
  TIME\_FIRST\_COMMENT &
  Time from first email/comment to current email/comment divided by the total time of the thread. &
  {[}0, 1{]} \\
 &
  CT1, CT2 &
  TIME\_TEXT\_LAST &
  Time from current email/comment to last email/comment divided by the total time of the thread. &
  {[}0, 1{]} \\
 &
  CT1, CT2 &
  TIME\_PREVIOUS\_COMMENT &
  Time from previous email/comment to current email/comment divided by the total time of the thread. &
  {[}0, 1{]} \\
 &
  CT1, CT2 &
  TIME\_TEXT\_NEXT &
  Time from current email/comment to next email/comment divided by total time of the thread. &
  {[}0, 1{]} \\ \bottomrule
\end{tabular}%
}
\footnotesize{\textit{Note: CT1 = classification task 1 on non-tone-bearing and tone-bearing emails/comments, CT2 = classification task 2 on civil and uncivil sentences.}}
\end{table*}

\textbf{\textit{Textual features:}}  We consider n-grams as textual features.
First, we perform text vectorization by transforming each word of the text into a feature, based on the absolute term frequencies. Second, we use 2-grams that represent the appearance of $n$ tokens sequences. Then, we use weighted TF-IDF to transform both features into numerical representations; \ie the frequency of words and 2-grams in the text are multiplied by their inverse document frequency~\cite{arya2019analysis}. We tune each model by using only the n-gram configuration that yields the best result.
	
\textbf{\textit{Conversational features}} describe the \textit{participants}, \textit{length}, \textit{structural}, and \textit{temporal} attributes of code review and issue discussions. Each one of these features is described in Table~\ref{table:features}, along with the classification tasks in which they are used. The conversational features include the following categories:

\begin{itemize}[leftmargin=*,nosep]

\item \textbf{Participant features} include features describing the discussion participants, \ie authors who wrote the code review email or issue comment as well as if the author is a reviewer or a developer. Participant features are relevant in our context because our previous work identified that emails/comments sent by maintainers exhibit different causes of incivility than those sent by other developers~\cite{ferreira2021shut}. Furthermore, maintainers send more uncivil code review emails than developers~\cite{ferreira2021shut}.

\item  \textbf{Length features} concern the length of emails, comments, or sentences. These specific features indicate length in terms of the number of characters in the email/comment or the sentence, as well as the relative number of words with respect to other emails, comments, or sentences. We included these length features because our previous work found that code review discussions with tone-bearing discussion features tend to be longer and have more uncivil than civil emails~\cite{ferreira2021shut}.

\item  \textbf{Structural features} describe the location of an email, comment, or sentence in relation to the entire email thread, issue thread, or the current email/comment itself. As shown by Ferreira \etal~\cite{ferreira2022how}, uncivil comments tend to emerge during the discussion in issue discussions, at various locations in the discussion thread. Hence, we hypothesize that structural features are pertinent to incivility classification.

\item  \textbf{Temporal features} concern the time that the email/comment was sent with respect to the immediately previous and next email/comment as well as the beginning and the end of the email/issue thread. Egelman \etal~\cite{egelman2020predicting} has shown that long reviewing time and long time working towards the solution of the problem are effective for flagging that code review authors have a feeling of pushback. Hence, we hypothesize that these features are suitable in our context because a long reply time might lead to incivility.

\end{itemize}

\subsection{Data Augmentation and Class Balancing}

Our datasets, especially the ones for civil and uncivil classification, are relatively small. To increase the training set and to boost performance for both classification tasks, we used the Easy Data Augmentation (EDA)~\cite{wei2019eda} techniques to augment the current datasets; the EDA techniques are known to contribute to performance gains of classifications when the dataset is small~\cite{wei2019eda}. Additionally, the datasets we use are highly imbalanced, skewing toward tone-bearing emails/comments and uncivil sentences. Machine learning classifiers are well known for underperforming when the data is skewed toward one class~\cite{japkowicz2002class, batista2004study}. To address this issue, we explored and evaluated three class balancing techniques that we describe in this section.

\subsubsection{Easy Data Augmentation Techniques (EDA)}
\label{section:eda}

In this study, we use the Easy Data Augmentation (EDA) Techniques~\cite{wei2019eda} to increase the size of our datasets. EDA is composed of four operations:

\begin{itemize}[leftmargin=*,nosep]
\item \textbf{Synonym Replacement (SR)} consists of randomly choosing $n$ words (excluding stop words) from the text and replacing them with a random synonym.

\item \textbf{Random Insertion (RI)} consists of finding a synonym of a random word in the text (excluding stop words) and inserting the synonym in a random position in the text. This process is repeated $n$ times.

\item \textbf{Random Swap (RS)} is when two words are randomly chosen and their positions are swapped. This is repeated $n$ times.

\item \textbf{Random Deletion (RD)} consists of randomly removing words in a sentence with probability $p$.
\end{itemize}

To find a synonym to perform the SR and RI operations, we use the NLTK \texttt{wordnet} corpus and the function \texttt{synsets (word)} to lookup for the word's synonym. Furthermore, we use the NLTK's list of English stopwords~\cite{nltk_stopwords} to exclude stopwords from the text. To mitigate the threat of long texts having more words than short texts, Wei and Zou~\cite{wei2019eda} suggest varying the number of words $n$ for SR, RI, and RS based on the text length $l$ with the formula $n = \alpha l$, where $\alpha$ is a hyperparameter that indicates the percentage of words in a text to be changed. Furthermore, for each original email/comment/sentence, it is possible to generate $n_{aug}$ augmented emails/comments/sentences. We evaluated different combinations of hyperparameters to augment the training set \CHANGED{(see our replication package for details)}. The hyperparameter search space was chosen based on the training set size and thresholds that have resulted in high performance in previous work by Wei and Zou~\cite{wei2019eda}. The hyperparameter tuning process is described in detail in Section~\ref{section:train_evaluate_classifiers}.

\subsubsection{Class Balancing Techniques}
\label{section:class_balancing_techniques}

To address the class imbalance problem of our dataset~\cite{japkowicz2002class}, we explored three class balancing techniques: \textit{random oversampling}, \textit{random undersampling}, and \textit{Synthetic Minority Oversampling Technique (SMOTE)}. We implemented these techniques using the Python library \texttt{imblearn} to compare their results when answering RQ1. The class balancing techniques are applied after the datasets are augmented by EDA.

\begin{itemize}[leftmargin=*,nosep]

    \item \textbf{Random oversampling} aims at taking random samples for the minority class and duplicating them until it reaches a size comparable to the majority class~\cite{padurariu2019dealing}.

    \item\textbf{Random undersampling} selects random samples from the majority class and removes them from the dataset until it reaches a size comparable to the minority class~\cite{batista2004study}.
        
    \item \textbf{SMOTE} is a method in which the minority class is oversampled by creating new samples and the majority class is undersampled~\cite{chawla2002smote}.
\end{itemize}

\subsection{Training and Evaluating the Classifiers}
\label{section:train_evaluate_classifiers}

\subsubsection{Classical Machine Learning Models}
\label{section:classifiers}

We consider six classical classifiers to detect incivility in code review and issues discussions (RQ1). The classifiers were implemented using the \texttt{sklearn} Python library.

\begin{itemize}[leftmargin=*,nosep]
    \item \textbf{Classification and Regression Tree (CART)} is a binary tree that aims at producing rules that predict the value of an outcome variable~\cite{lewis2000introduction}.

    \item \textbf{k-nearest neighbors (KNN)} assumes that similar datapoints are close to each other. Hence, the algorithm relies on distance metrics for classification. The resulting class is the one that has the nearest neighbors~\cite{shah2020comparative}.
        
    \item \textbf{Logistic Regression (LR)} uses a logistic function to model the dependent variable. The goal of the algorithm is to find the best fitting model to describe the relationship between the dependent and independent variable~\cite{shah2020comparative}.
    
    \item\textbf{Naive Bayes (NB)} is a probabilistic classifier based on the Bayes theorem. It assumes that the presence (or absence) of a particular feature of a class is unrelated to the presence (or absence) of any other feature~\cite{dai2007transferring}. In this work, our text classification is performed using the Multinomial Naive Bayes model that has improved performance over the Bernoulli model for text classification~\cite{rennie2003tackling}.
            
    \item \textbf{Random Forest (RF)} is a group of decision trees whose nodes are defined based on the training data~\cite{shah2020comparative}. The most frequent label found by the trees from the forest is the resulting class~\cite{arya2019analysis}.
    
    \item \textbf{Support Vector Machine (SVM)} is a linear model that creates a hyperplane separating the data into two classes~\cite{goudjil2018novel}.
\end{itemize}

During the training process, we performed \textit{nested cross-validation} with \textit{grid search}~\cite{bergstra2011algorithms} to test a combination of hyperparameters and evaluate the models' performance. Specifically, we first split the dataset into train and test sets in the outer stratified 5-fold cross-validation for model evaluation. The training set obtained from the outer cross-validation is then further split into training (for training the models) and validation (for selecting the best hyperparameters) sets in the inner stratified 5-fold cross-validation. \CHANGED{The search space for the hyperparameters of each model, which were defined according to the literature, can be found in our replication package.} In each outer cross-validation fold, we used the performance metrics presented in Section~\ref{section:performance_metrics} to evaluate the performance of the classifier.

\subsubsection{BERT Base Model}
\label{section:bert}

We use the uncased BERT base model, pretrained on the English language~\cite{bert-base-uncased-hugging-face}, to detect incivility. We chose to use the uncased model (\ie the model does not make a difference between ``english'' and ``English'') because the case information is not relevant to our classification tasks. Furthermore, due to the large number of parameters in this model (approximately 110 million parameters), we did not pretrain it from scratch to reduce the risk of overfitting~\cite{turc2019}. The classification task is done using the \texttt{Transformers} PyTorch library with \texttt{AutoModelFor\-SequenceClassification}~\cite{auto_class_sequence_classification}, which has a classification head. 

We split the input dataset into train, test, and evaluation datasets in a 70-15-15 ratio stratified along the labels. To optimize BERT's hyperparameters, we run \textit{bayesian optimization}~\cite{snoek2012practical} with 50 trials for each one of the EDA parameter settings (see Section~\ref{section:eda}), \ie eight times. BERT's hyperparameter optimization was done using the \texttt{hyperparameter\_search()} function~\cite{hyperparameter_search} from the \texttt{Trainer} class with \texttt{optuna} as the backend. The Bayesian optimization takes in the training and evaluation sets as inputs; the former is used to train the model with different hyperparameters and the latter is used to select the best hyperparameters. \CHANGED{The search space for the hyperparameters can be found in our replication package.}

After obtaining the best set of hyperparameters, we perform a 5-fold cross-validation to train and test BERT. For that, we use the \texttt{Trainer}~\cite{trainer_class} class from the \texttt{Transformers} library. The training adopts an epoch evaluating strategy, \ie evaluating BERT's performance at the end of each epoch using the performance metrics described in Section~\ref{section:performance_metrics}.

\subsubsection{Performance metrics}
\label{section:performance_metrics}

To compare the performance of our classifiers, we evaluate their performances using the confusion matrix: TP is the number of true positives, FN is the number of false negatives, FP is the number of positives, and TN is the number of true negatives.

Based on this matrix, we first computed two well-known metrics, namely \textit{precision} and \textit{recall}~\cite{baeza1999modern}. The \textbf{precision} of a given target class (\ie non-tone-bearing, tone-bearing, civil, or uncivil) is defined by the ratio of data points (emails, comments, or sentences) for which a given classifier correctly predicted the target class; \ie $ precision = TP/(TP + FP)$. The precision value is always between 0 (worst possible score) and 1 (perfect score). In each classification task of our experiments, we first calculated the precision in each class, then the macro-average metric across both classes to represent the overall precision.

The \textbf{recall} of a given target class is the ratio of all data points with the target class that a given classifier was able to correctly find; \ie $ recall = TP/(TP + FN)$. The recall value is also between 0 (worst possible score) and 1 (perfect score). Similar to precision, we calculated per-class and macro-averaged recall metrics for each classification task.

Then, to have a single value representing the goodness of the models, we computed the \textbf{F1-score}, which is the harmonic mean of precision and recall, \ie $ F1 = 2 \cdot\frac{precision \cdot recall}{precision + recall}$. The F1-score is independent of the number of true negatives and it is highly influenced by classes labeled as positive. The F1-score is always between 0 (lowest precision and lowest recall) and 1 (highest precision and highest recall). In our experiments, we calculated per-class and macro-averaged F1 metrics.

Finally, we computed the \textbf{Matthews Correlation Coefficient (MCC)}~\cite{matthews1975comparison}, which is a single-value classification metric that is more interpretable and robust to changes in the prediction goal~\cite{chicco2020advantages}, because it summarizes the results of all four quadrants of a confusion matrix~\cite{croft2021empirical}. The MCC metric is calculated as
$ MCC = \frac{TP \cdot TN - FP \cdot FN}{\sqrt{(TP+FP) (TP+FN)(TN+FP)+(TN+FN)}}$ and its value is always between -1 (worst possible score) and 1 (perfect score), with 0 suggesting that the model's performance is equal to random prediction. To calculate TP, TN, FP, and FN for MCC, we considered civil and non-tone-bearing as positive classes and uncivil and tone-bearing as negative classes, although this selection does not influence the end results. To be able to compare the MCC scores with the other performance metrics (\ie macro-averaged precision, recall, and F1-score), we normalized the MCC values to the $[0,1]$ interval. Therefore, \textbf{the normalized MCC (nMCC)} is defined by $nMCC = \frac{MCC + 1}{2}$~\cite{chicco2020advantages, chicco2021benefits}. We also use nMCC as the primary metric during hyperparameter evaluation (Section~\ref{section:classifiers}). 

\subsection{Experimental Design to Answer the RQs}
\label{sec:methodology:rqs}

\subsubsection{Detecting Incivility (RQ1)}
\label{sec:methodology:rq1}

In RQ1, we have two classification tasks for each dataset, \ie (1) classification of code review emails and issue comments into non-tone-bearing and tone-bearing and (2) classification of code review sentences and issue sentences into civil and uncivil. 

For each classification task and for each dataset, we compare BERT with six classical machine learning models. We assess BERT with three class balancing conditions: no class balancing, random oversampling, and random undersampling. It was not possible to run SMOTE with BERT because the current SMOTE implementations need to convert textual features to numerical vectors (via tokenization to a form suitable for SMOTE)~\cite{he_ma_2013} and cannot be applied to textual features that are used for BERT. The classical machine learning models are assessed with three class balancing techniques: random oversampling, random undersampling, and SMOTE. Thus, for each classification task and each dataset, we have 21 experimental conditions (7 classifiers * 3 balancing techniques). The hyperparameters are tuned separately for each combination of classification task, dataset, classifier, and class balancing technique. 
In this paper, we report the results related to hyperparameters that had the best-averaged nMCC score across all outer folds.

\CHANGED{\subsubsection{Adding the Previous Email or Comment When Detecting Incivility (RQ2)}
\label{sec:methodology_rq2}

To answer RQ2, we add the previous email/comment from the same email or issue thread and concatenate it with the original email/comment to create a new dataset. For this RQ, we only focus on BERT since it provided the best performance in the initial detection. We use the hyperparameters, the EDA parameter configuration, and the imbalance handling technique that obtained the best nMCC score in RQ1 for BERT in each classification task for each dataset. To analyze if adding the previous email/comment helps to detect incivility, we compare the difference between the performance scores of RQ1 and RQ2. Hence, for each performance metric (PM), $\Delta PM = PM_{RQ2} - PM_{RQ1}$. We consider that the previous email/comment does not help if $\Delta PM<0$ and that it helps if $\Delta PM>0$.}

\subsubsection{Detecting Incivility in a Cross-Platform Setting (RQ3)}

To answer RQ3, we train our models and test them in the other dataset for each classification task; \ie we train our classifiers on the code review dataset and test them on the GitHub issues dataset, and vice versa. For that, we use the hyperparameters, the EDA parameters configuration, and the imbalance handling techniques that obtained the best nMCC score in RQ1 for the dataset used to train the classifiers. Because the best hyperparameters can differ among the five folds for the classical machine learning models, we pick the hyperparameters that were chosen most frequently across all five folds. If there is a tie between two sets of hyperparameters, we then choose the hyperparameter from the fold that had the highest nMCC score.

\CHANGED{\subsubsection{Analyzing misclassified cases (RQ4)}
To answer RQ4, we assess the tone-bearing discussion features (TBDFs)~\cite{ferreira2021shut} that the analyzed models misclassified. The misclassified sentences were extracted from the test sets in the outer fold cross-validation (see Section~\ref{section:classifiers}). Since the non-tone-bearing emails/comments are split into sentences and the sentence classification (CT2) depends on the tone that they demonstrate~\cite{ferreira2021shut}, we calculate the percentage of sentences that were misclassified by each classifier per TBDF in CT2.}

\section{Results}

In this section, we present the results for each research question for the non-tone-bearing/tone-bearing and civil/uncivil classification tasks. We only report the results of the hyperparameters that yielded the best results.

\subsection{Models' Performance on Incivility Detection (RQ1)}

\subsubsection{Classification into Non-tone-bearing and Tone-bearing}

\begin{figure*}[ht!]
\centering
\includegraphics[clip,trim=0cm 3.5cm 0cm 3.5cm, width=0.9\linewidth]{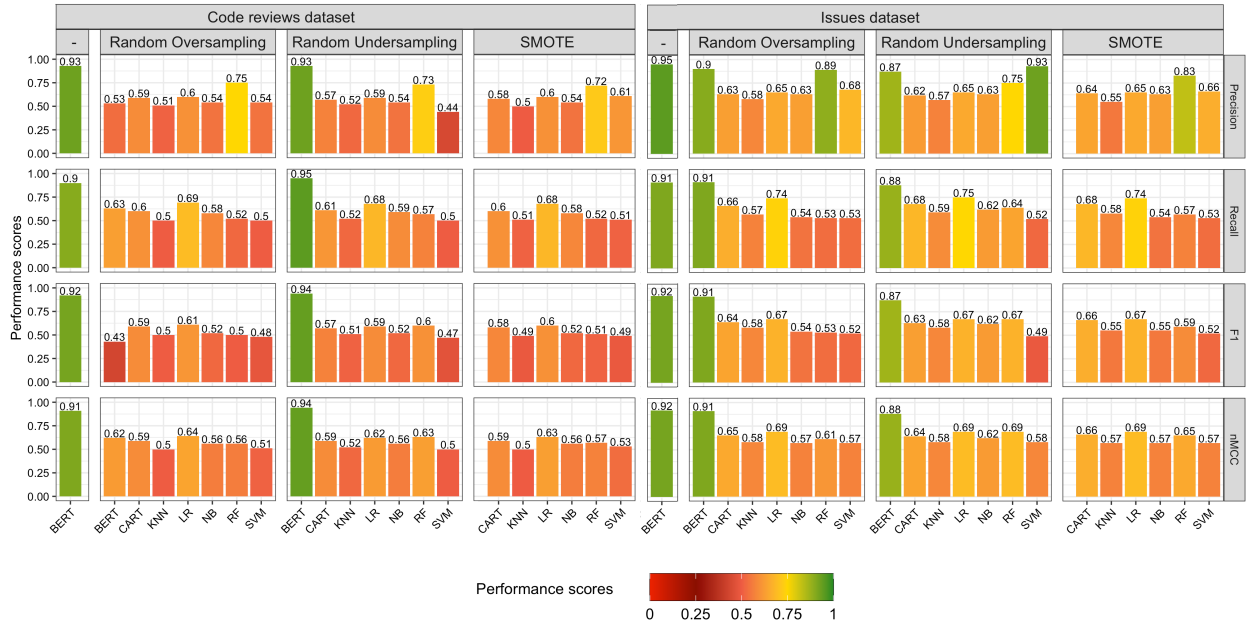}
\caption{Average performance scores per class balancing technique and classifier for the classification of non-tone-bearing and tone-bearing emails/comments (CT1).}
\label{fig:rq1_tech_non_tech}
\end{figure*}

\begin{figure*}[ht!]
\centering
\includegraphics[clip,trim=0cm 3.5cm 0cm 3.5cm, width=0.9\linewidth]{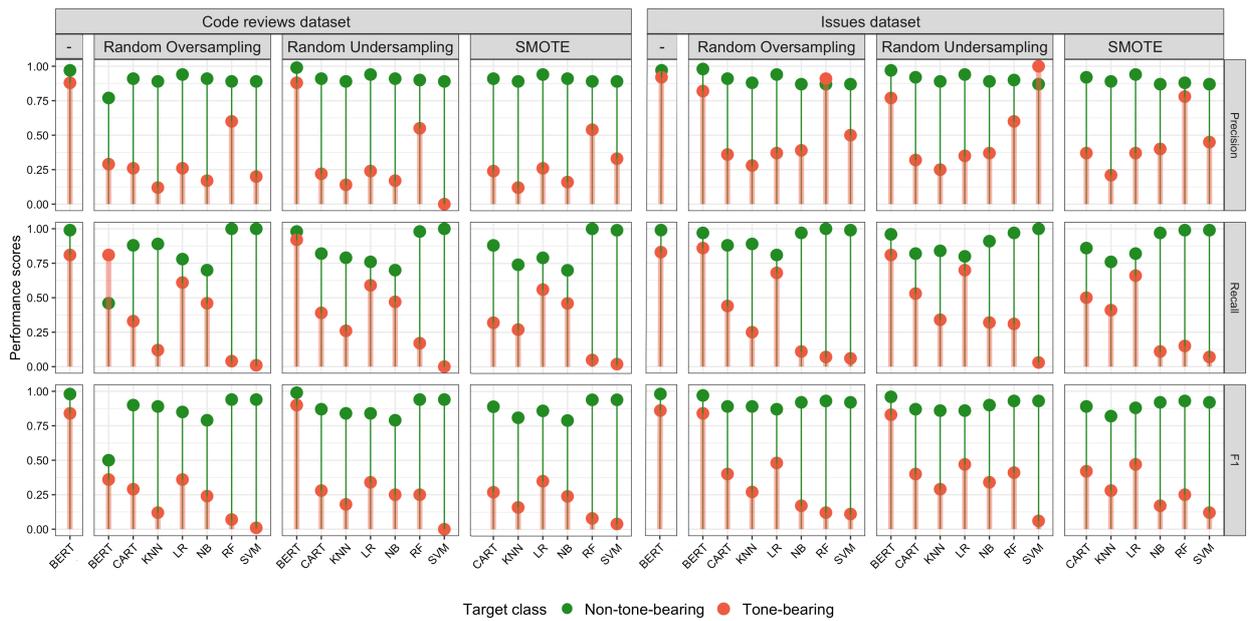}
\caption{Performance scores per target class for the classification of non-tone-bearing and tone-bearing emails/comments (CT1).}
\label{fig:rq1_tech_non_tech_per_class}
\end{figure*}

Figure~\ref{fig:rq1_tech_non_tech} presents the average performance scores for each experiment condition for the code reviews (left) and issues (right) datasets and Figure~\ref{fig:rq1_tech_non_tech_per_class} shows the performance per target class.

\textbf{For the code reviews dataset, BERT \textit{without class balancing} and \textit{with random undersampling} has the best performance compared to the classical classifiers}, with $F1=0.92$ and $0.94$, and $nMCC = 0.91$ and $0.94$, respectively. \textbf{Classical machine learning models underperform to classify tone-bearing code review emails.} The nMCC scores for the classical classifiers and for BERT \textit{with random oversampling} (ranging from 0.50 to 0.64) are very low compared to BERT's nMCC scores \textit{with random undersampling} and \textit{without class balancing} (0.94 and 0.91, respectively), showing that such classifiers are not effective to detect tone-bearing code review emails. Figure~\ref{fig:rq1_tech_non_tech_per_class} (left) confirms this result, indicating that the tone-bearing class (red color) having overall lower precision and recall than the non-tone-bearing class (green color) for the underperforming classifiers. We also observe that among the classical classifiers, Random Forest (RF) achieved the highest precision ($\approx$0.7) regardless of the class balancing technique, but with a low recall ($\approx$0.5); on the contrary, Logistic Regression (LR) achieved the highest recall ($\approx$0.7), but a relatively low precision ($\approx$0.6).

\textbf{For the issues dataset, BERT also performs better than the classical classifiers in all class balancing conditions.} Similar to the code reviews dataset, BERT is able to precisely classify non-tone-bearing and tone-bearing issue comments ($precision\approx0.9$), finding a substantial number of issue comments ($recall\approx0.9$), and effectively classifying the non-tone-bearing and tone-bearing issue comments  ($nMCC\approx0.9$), in all class balancing conditions. Furthermore, \textbf{classical classifiers also underperform to classify tone-bearing issue comments.} The nMCC scores range from 0.57 to 0.69 demonstrating that tone-bearing issue comments are not effectively detected with classical machine learning models (see Figure~\ref{fig:rq1_tech_non_tech_per_class} (right)), except for RF with \textit{random oversampling} and SVM with \textit{random undersampling}, in which their precision metrics are better for the tone-bearing class ($precision_{RF}=0.91$ and $precision_{SVM}=1.0$) than the non-tone-bearing class ($precision_{RF}=0.87$ and $precision_{SVM}=0.87$); yet their recalls are very low for the tone-bearing class ($recall_{RF}=0.07$, $recall_{SVM}=0.03$).

It is surprising that \textbf{BERT has a good performance overall even without any class balancing technique.} This result is confirmed by Figure~\ref{fig:rq1_tech_non_tech_per_class}, which shows that even without a class balancing technique both non-tone-bearing and tone-bearing classes have a good F1-score for both the code reviews dataset ($tone-bearing = 0.84$ and $non-tone-bearing = 0.98$) and the issues dataset ($tone-bearing = 0.86$ and $non-tone-bearing = 0.98$).

\begin{figure*}[ht!]
\centering
\includegraphics[clip,trim=0cm 3.5cm 0cm 3.5cm, width=0.9\linewidth]{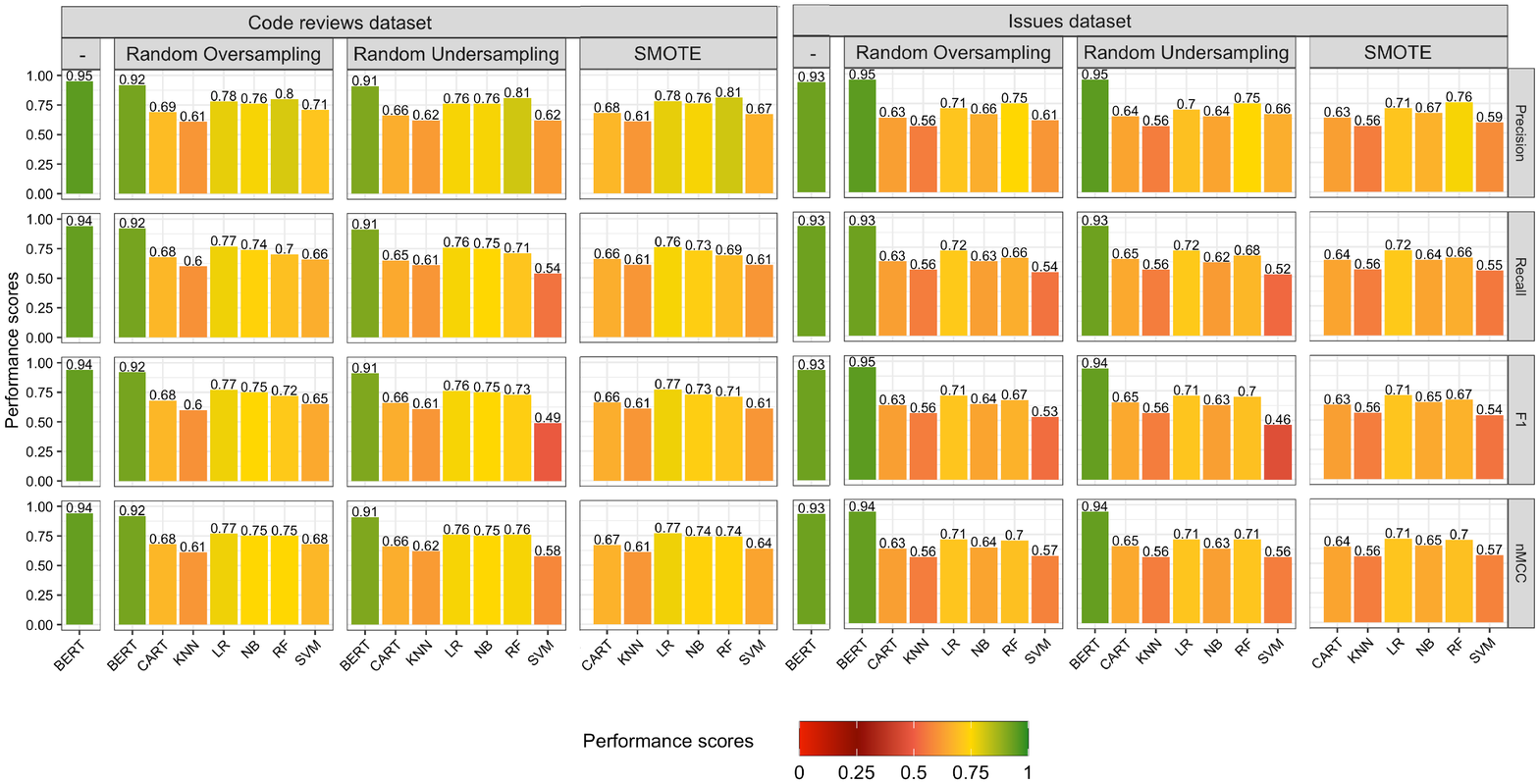}
\caption{Average performance scores per class balancing technique and classifier for the classification of civil and uncivil sentences (CT2).}
\label{fig:rq1_civil_uncivil}
\end{figure*}

\begin{figure*}[ht!]
\centering
\includegraphics[clip,trim=0cm 3.5cm 0cm 3.5cm, width=0.9\linewidth]{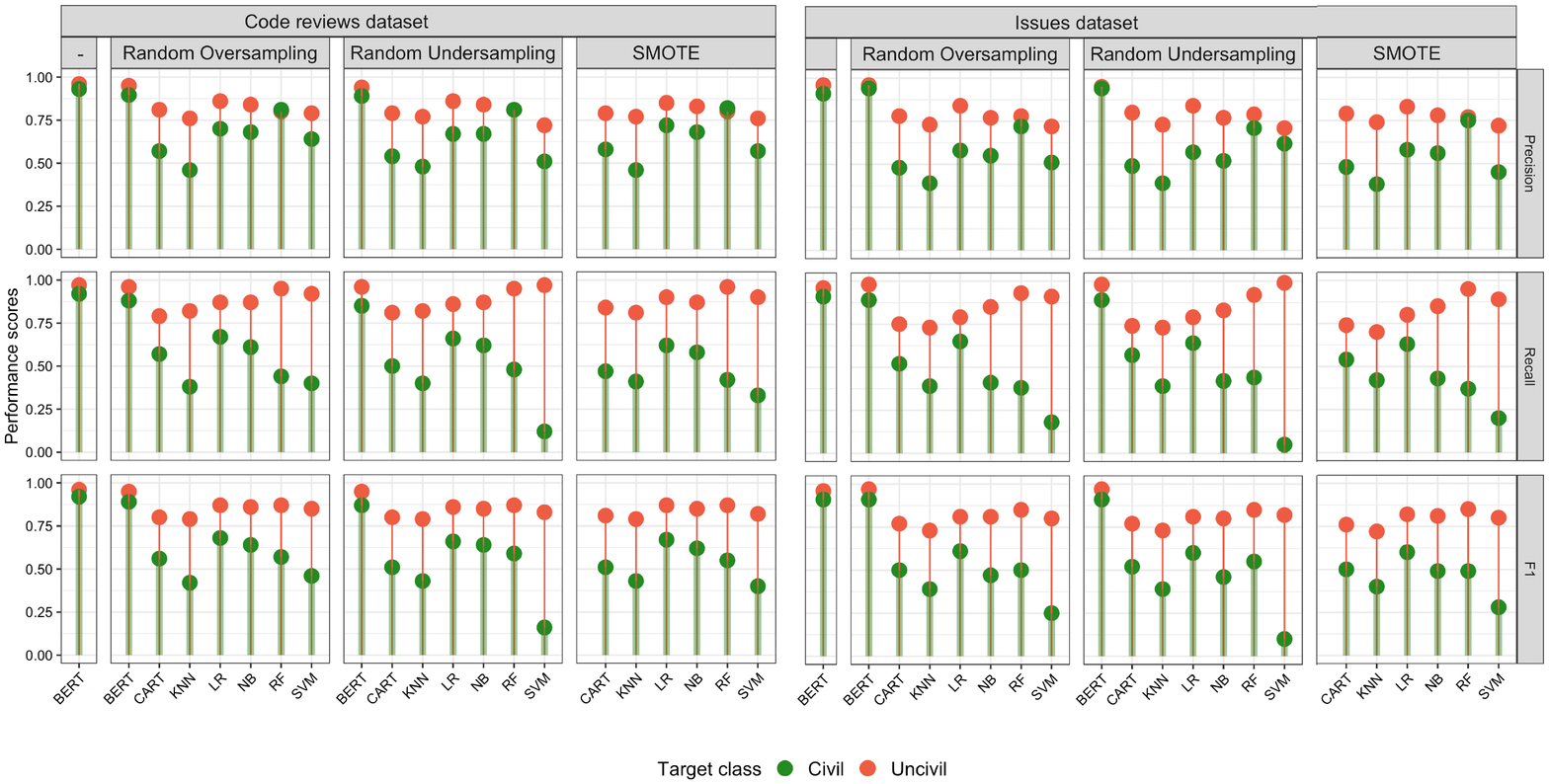}
\caption{Performance scores per target class for the classification of civil and uncivil sentences (CT2).}
\label{fig:rq1_civil_uncivil_per_class}
\end{figure*}

\subsubsection{Classification into Civil and Uncivil}

Figure~\ref{fig:rq1_civil_uncivil} illustrates the performance metrics for each experiment setting and for both the code reviews (left) and issues (right) datasets. Similarly, Figure~\ref{fig:rq1_civil_uncivil_per_class} presents the performance metrics per target class for both datasets.

\textbf{BERT is the best performing classifier regardless of the class balancing technique for the code reviews dataset.} We observe that, similar to the non-tone-bearing/tone-bearing classification task, BERT has the best performance ($precision\approx0.9$, $recall\approx0.9$, $F1\approx0.9$, $nMCC\approx0.9$) to classify civil and uncivil code review sentences. However, classical machine learning techniques tend to perform better in the classification of civil/uncivil sentences than in non-tone-bearing/tone-bearing emails, 
with nMCC scores ranging from 0.58 to 0.77 (compared to between 0.50 and 0.64 for the non-tone-bearing/tone-bearing classification). Additionally, \textbf{the classical models underperform when classifying civil code review sentences} (see Figure~\ref{fig:rq1_civil_uncivil_per_class} (left)). The Logistic Regression, Naive Bayes, and Random Forest classifiers have overall promising results though, with $precision\approx0.8$ and $recall\approx0.7$. 

\textbf{For the issue comments dataset, BERT is also the best classifier for detecting incivility regardless of which class balancing technique is used} ($precision\approx0.9$, $recall\approx0.9$). Although classical techniques also tend to underperform when classifying civil issue sentences (see Figure~\ref{fig:rq1_civil_uncivil_per_class} (right)), Logistic Regression and Random Forest have good precision ($\approx0.71$ for LR and $\approx0.75$ for RF) and recall ($\approx0.72$ for LR and $\approx0.67$ for RF) overall.

\begin{mdframed}[backgroundcolor=black!10,rightline=true,leftline=true]
\textbf{Summary RQ1:} BERT performs better than the classical machine learning classifiers regardless of the class balancing technique for non-tone-bearing/tone-bearing and civil/uncivil classification in both datasets. Classical machine learning techniques tend to underperform when classifying the tone-bearing and civil classes.
\end{mdframed}

\subsection{\CHANGED{Adding the Previous Email or Comment in Incivility Detection (RQ2)}}

\begin{figure*}[ht!]
\centering
\subfloat[Non-tone-bearing and tone-bearing classification (CT1)]{\includegraphics[clip,trim=0cm 0.5cm 0cm 0cm,width=0.49\linewidth]{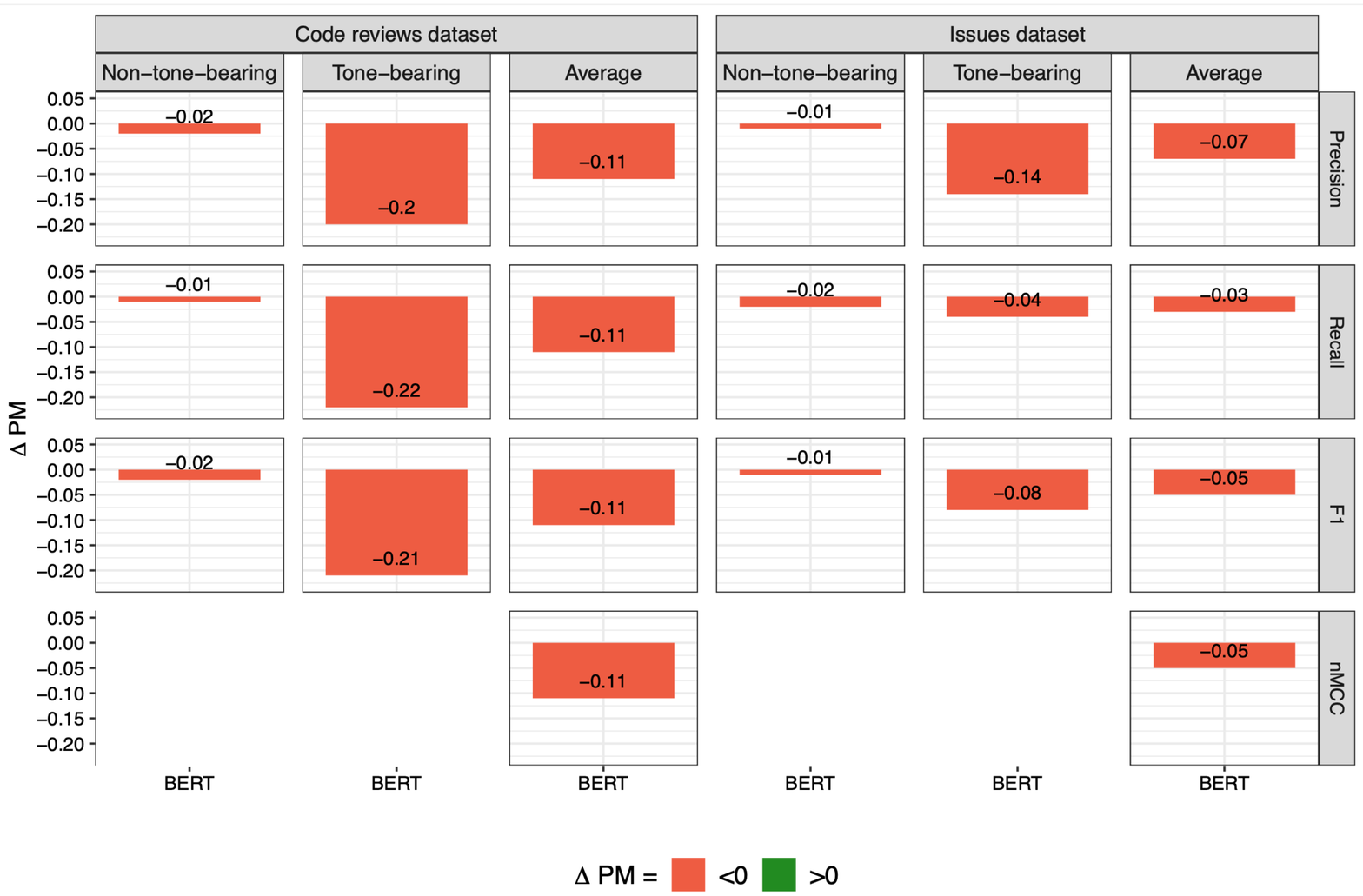}}
\hfill
\subfloat[Civil and uncivil classification (CT2)]{\includegraphics[clip,trim=0.5cm 1cm 0.5cm 1cm, width=0.49\linewidth]{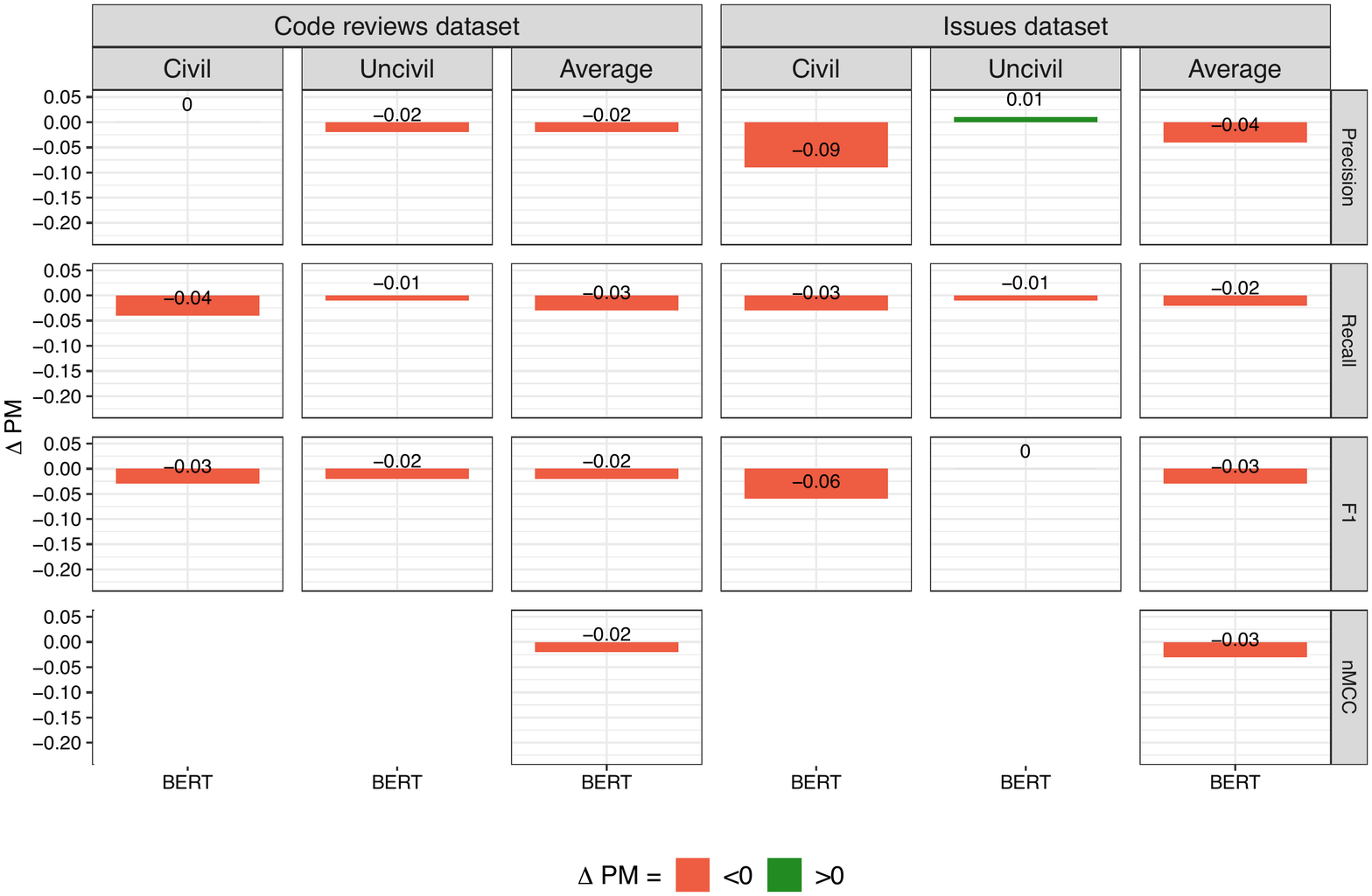}}
\caption{Difference of BERT's performance scores between RQ1 (without the previous email/comment) and RQ2 (with the previous email/comment).}
\label{fig:context_difference}
\end{figure*}

Figure~\ref{fig:context_difference} (a) and (b) present the difference in performance metrics between the BERT results \CHANGED{considering the previous email or comment (RQ2) and without (RQ1)}, for the two classifications respectively.

We found that \textbf{\CHANGED{adding the previous email/comment} does not help to classify Non-tone-bearing and tone-bearing code review emails and issue comments.} We observe that, for both datasets when detecting Non-tone-bearing and tone-bearing contents, $\Delta PM$ is negative overall, with the tone-bearing class results having most drastically decreased \CHANGED{when the previous email or comment is added} (Figure~\ref{fig:context_difference} (a)). BERT's performance on the code reviews dataset, more specifically, gets worse by $\approx-0.2$ for the tone-bearing class, having its precision decreased from 0.88 to 0.67 and its recall from 0.92 to 0.71. Similarly, on the issues dataset, BERT's precision for tone-bearing comments decreased by 0.14, going from 0.92 to 0.78; and the recall decreased by 0.04, from 0.83 to 0.79.

\textbf{Overall, \CHANGED{adding the previous email/comment} also does not help to classify civil and uncivil code review and issue sentences.} Our results show that the civil class results have significantly decreased, especially for the issues dataset (Figure~\ref{fig:context_difference} (b)). Although the precision did not change for the code reviews dataset, its recall decreased by 0.04, going from 0.92 to 0.88). For the issues dataset, the precision and recall decreased by 0.09 and 0.03, respectively.

\begin{mdframed}[backgroundcolor=black!10,rightline=true,leftline=true]
\textbf{Summary RQ2:} Adding the previous code review email and issue comment makes the prediction worse for both non-tone-bearing/tone-bearing and civil/uncivil classification. The effect is stronger for the tone-bearing class.
\end{mdframed}
\subsection{Incivility Detection in a Cross-Platform Setting (RQ3)}


\subsubsection{Classification into Non-tone-bearing and Tone-bearing}

\begin{figure*}[ht!]
\centering
\subfloat[Non-tone-bearing and tone-bearing classification (CT1)]{\includegraphics[clip,trim=0.5cm 4cm 0.5cm 4cm, width=0.49\linewidth]{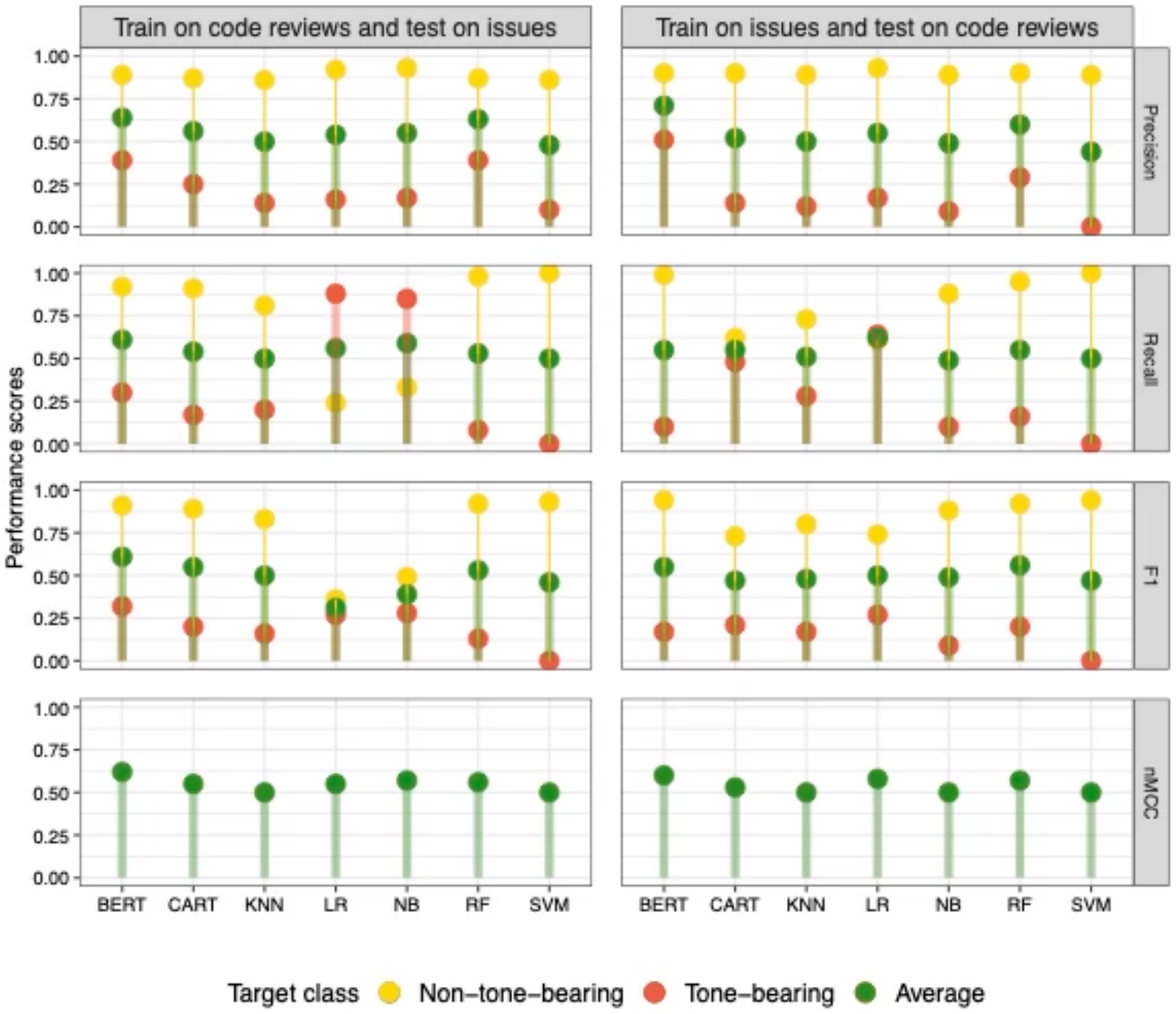}}
\hfill
\subfloat[Civil and uncivil classification (CT2)]{\includegraphics[clip,trim=0.5cm 4cm 0.5cm 4cm, width=0.49\linewidth]{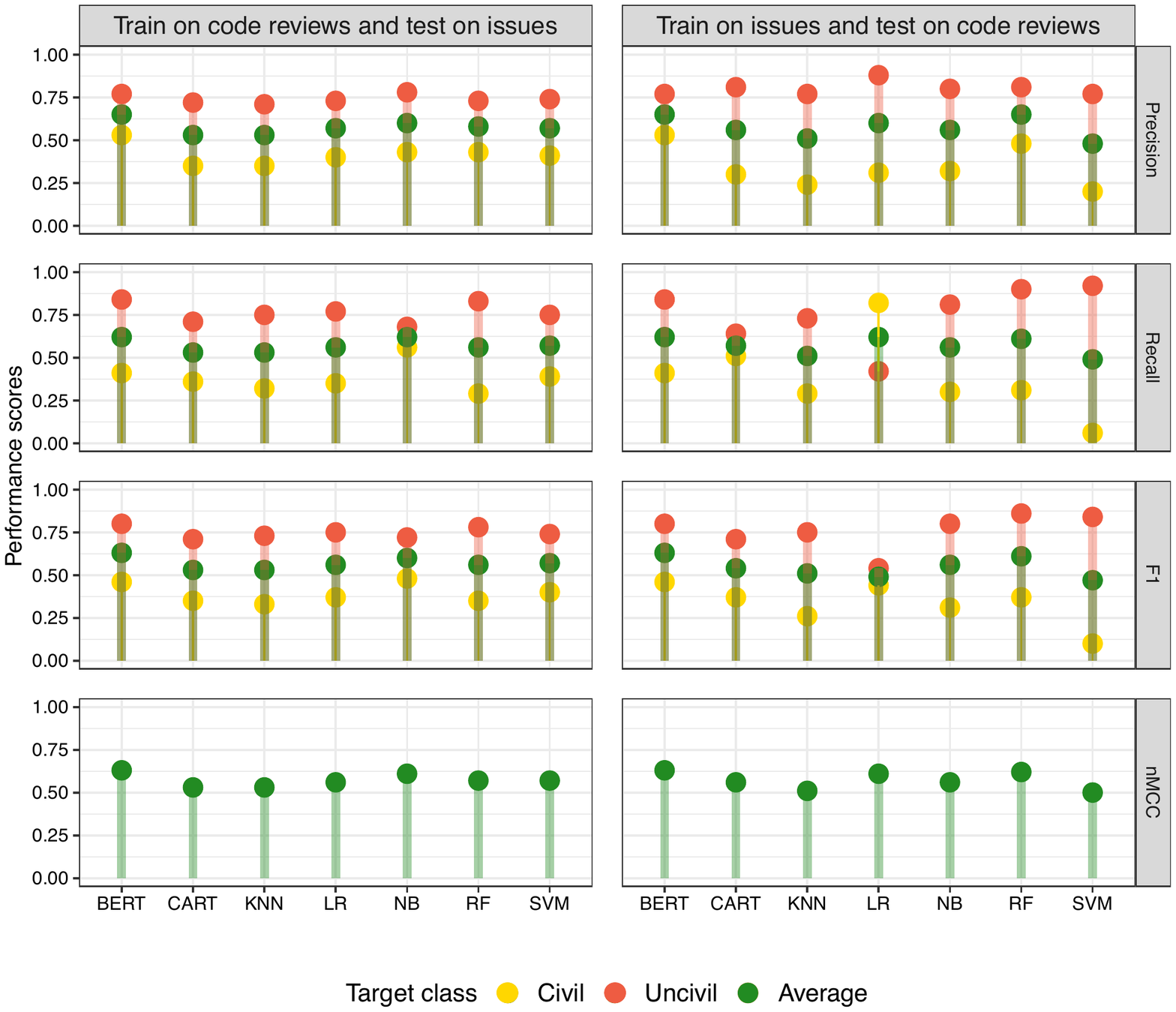}}
\caption{Performance scores for classification in a cross-platform setting.}
\label{fig:cross_platform_both}
\end{figure*}

Figure~\ref{fig:cross_platform_both} (a) presents the performance for non-tone-bearing and tone-bearing classification in the cross-platform setting. \textbf{The classifiers' performances degraded to classify tone-bearing} discussions in a cross-platform setting.  When training our classifiers on the code reviews dataset and testing them on the issues dataset, we observe that BERT is the best classifier, with a nMCC score of 0.62. Our results show that the classifiers' performances are not satisfactory in precisely classifying tone-bearing discussions (red color), with precision scores ranging from 0.10 (SVM) to 0.39 (BERT and RF). Interestingly, the Logistic Regression (LR) and Naive Bayes (NB) classifiers can retrieve a significant percentage of tone-bearing discussions ($recall_{LR} = 0.88$, $recall_{NB} = 0.85$); even though they fail to precisely classify such cases ($precision_{LR} = 0.16$,  $precision_{NB} = 0.17$). We also observe a similar pattern when training on the issues dataset and testing on the code reviews dataset. The MCC scores ranged from 0.50 (KNN and SVM) to 0.62 (BERT). In this setting, BERT is more precise ($precision = 0.51$) than in the previous setting ($precision=0.39$) at classifying tone-bearing discussions, yet the coverage is lower ($recall = 0.10$ in this setting, versus original $recall = 0.30$). Surprisingly, the Logistic Regression classifier has a similar recall for both target classes ($recall_{non-tone-bearing} = 0.61$, $recall_{tone-bearing} = 0.64$).

\subsubsection{Classification into Civil and Uncivil}

Figure~\ref{fig:cross_platform_both} (b) presents the results for the civil and uncivil classification in the cross-platform setting. \textbf{The classifiers' performances are also degraded to classify civil sentences in a cross-platform setting.} When training on code reviews and testing on issues, we observe that all classifiers are able to precisely classify uncivil discussions with $precision \approx 0.7$ with good coverage ($recall \approx 0.8$). However, all classifiers have low precision (ranging from 0.35 to 0.53) and low recall (ranging from 0.29 to 0.56). When training on issues and testing on code reviews, we observe the same pattern as in the aforementioned configuration, \ie all classifiers can precisely classify the uncivil class ($precision \approx 0.8$) with a $recall \approx 0.7$. Interestingly, the Logistic Regression classifier has a recall higher for the civil class ($recall=0.82$) than the uncivil class ($recall=0.42$).

\begin{mdframed}[backgroundcolor=black!10,rightline=true,leftline=true]
\textbf{Summary RQ3:} None of the classifiers are effective to classify tone-bearing and civil discussions in a cross-platform setting. However, all classifiers were able to perform well when classifying the non-tone-bearing and uncivil classes in a cross-platform setting.
\end{mdframed}
\CHANGED{\subsection{Analysis of misclassified cases (RQ4)}
\label{sec:discussion_misclassification}

\subsubsection{Misclassified TBDFs per incivility classifier}
\label{sec:discussion_rq1}

\textbf{Contrary to the classical machine learning models, BERT can correctly classify more than 70\% of the sentences for \textit{all} civil and uncivil TBDFs for the code reviews and issues datasets.} As Figure~\ref{fig:discussion_rq1} shows, \textbf{in the code reviews dataset BERT mostly misclassifies sentences demonstrating the civil TBDFs \textit{friendly joke} (28.57\%), \textit{commanding} (22.22\%), and \textit{sadness} (14.29\%) and the uncivil TBDFs \textit{irony} (16.67\%), \textit{threat} (15.28\%), and \textit{vulgarity} (15.28\%)}. Although BERT misses 22.22\% of sentences expressing \textit{commanding}, the classical machine learning models are worst in classifying this TBDF (varying from 44.44\% for LR and SVM to 88.89\% for KNN). Concerning the \textit{friendly joke} TBDF, although SVM is as good as BERT (both models miss 28.57\% of sentences with this TBDF), the other machine learning models misclassify from 42.86\% (for CART, LR, and NB) to 71.43\% (for KNN AND RF) sentences. The same happens for \textit{commanding} and \textit{sadness}. Interestingly, \textbf{the classical machine learning models perform better than BERT to identify the uncivil TBDFs that BERT misses in the code reviews dataset}. That is, KNN and SVM misclassify 14.29\% of sentences (instead of 16.67\% for BERT) demonstrating \textit{irony}; CART and KNN misclassify 12.50\% of sentences (instead of 15.38\% for BERT) showing \textit{threat}; CART, LR, and SVM do not miss any sentence demonstrating \textit{vulgarity}, while BERT misses 15.38\%.

Note that, as demonstrated by Ferreira \etal, the TBDFs named \textit{confusion}, \textit{criticizing oppression}, \textit{dissatisfaction}, and \textit{expectation} were only encountered in issue discussions~\cite{ferreira2022how} and not in code review discussions~\cite{ferreira2021shut}. \textbf{For the issues dataset, BERT mainly misclassifies the following civil TBDFs: \textit{friendly joke} (26.09\%), \textit{criticizing oppression} (17.65\%), \textit{considerateness} (13.56\%), and \textit{dissatisfaction} (12.90\%). None of the uncivil TBDFs had more than 10\% of misclassified sentences.} Furthermore, none of the classical machine learning models can classify the aforementioned TBDFs better than BERT.

\begin{figure}[ht!]
\centering
\includegraphics[clip, width=\linewidth]{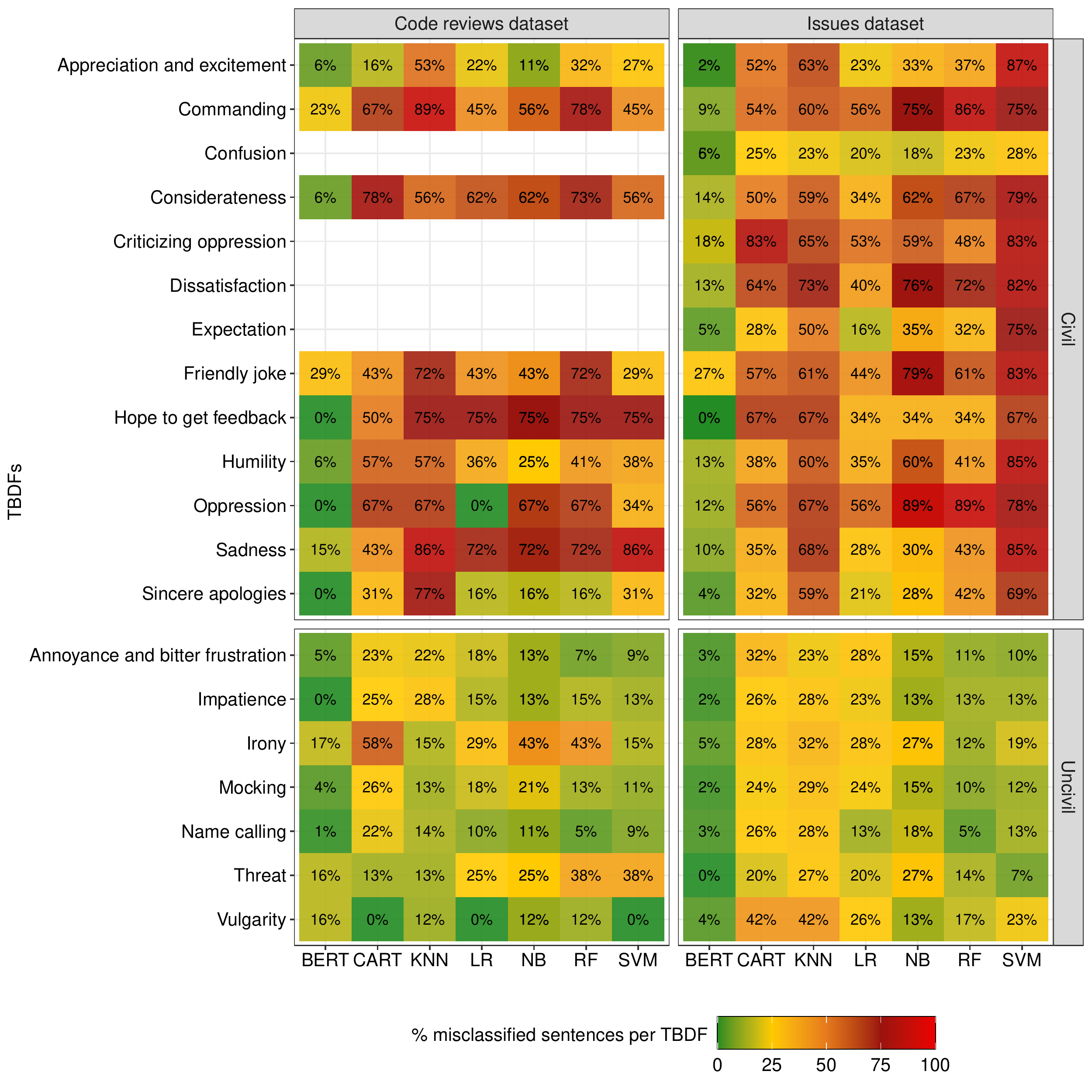}
\caption{\CHANGED{Percentage of misclassified sentences per TBDF per classifier.}}
\label{fig:discussion_rq1}
\end{figure}

\begin{mdframed}[backgroundcolor=black!10,rightline=true,leftline=true]
\textbf{Summary RQ4 (part 1):} \textit{Irony} and \textit{vulgarity} are are among the most difficult uncivil TBDFs to classify by BERT, while \textit{friendly joke} and \textit{commanding} are the most difficult uncivil TBDFs.
\end{mdframed}

\subsubsection{BERT's misclassified TBDFs considering the previous email or comment}

Figure~\ref{fig:discussion_rq2} presents the percentage of misclassified sentences per TBDF for BERT considering the previous email or comment. Surprisingly, for the code reviews dataset, \textit{commanding}, \textit{friendly joke}, \textit{sadness}, and \textit{threat} that were most frequently misclassified by BERT without the previous email or comment (see Section~\ref{sec:discussion_rq1}) now have 100\% of the sentences correctly classified. \textit{Irony} has a slightly decreased number of misclassified sentences, going from 16.67\% to 14.29\%, and \textit{vulgarity} has an increased number of misclassified sentences, by 9.62\% (from 15.38\% to 25\%). Additionally, \textit{appreciation and excitement}, \textit{sincere apologies}, \textit{impatience}, and \textit{mocking} were more frequently misclassified considering the previous email or comment than without, increasing the number of misclassified sentences by 9.82\%, 11.11\%, 7.14\%, and 7.98\%, respectively.

For the issues dataset, the number of misclassified sentences was decreased by 16.09\% for the \textit{friendly joke} TBDF and by 1.67\% for the \textit{considerateness} TBDF. \textit{Criticizing oppression} has an increased number of misclassified sentences by 11.76\%, and \textit{dissatisfaction} by 0.29\%. Furthermore, \textit{expectation}, \textit{sadness}, and \textit{sincere apologies} were the most impacted TBDFs, increasing the number of misclassified sentences by 8.38\%, 22.62\%, and 14.16\%, respectively.

\begin{figure}[t]
\centering
\includegraphics[clip, width=\linewidth]{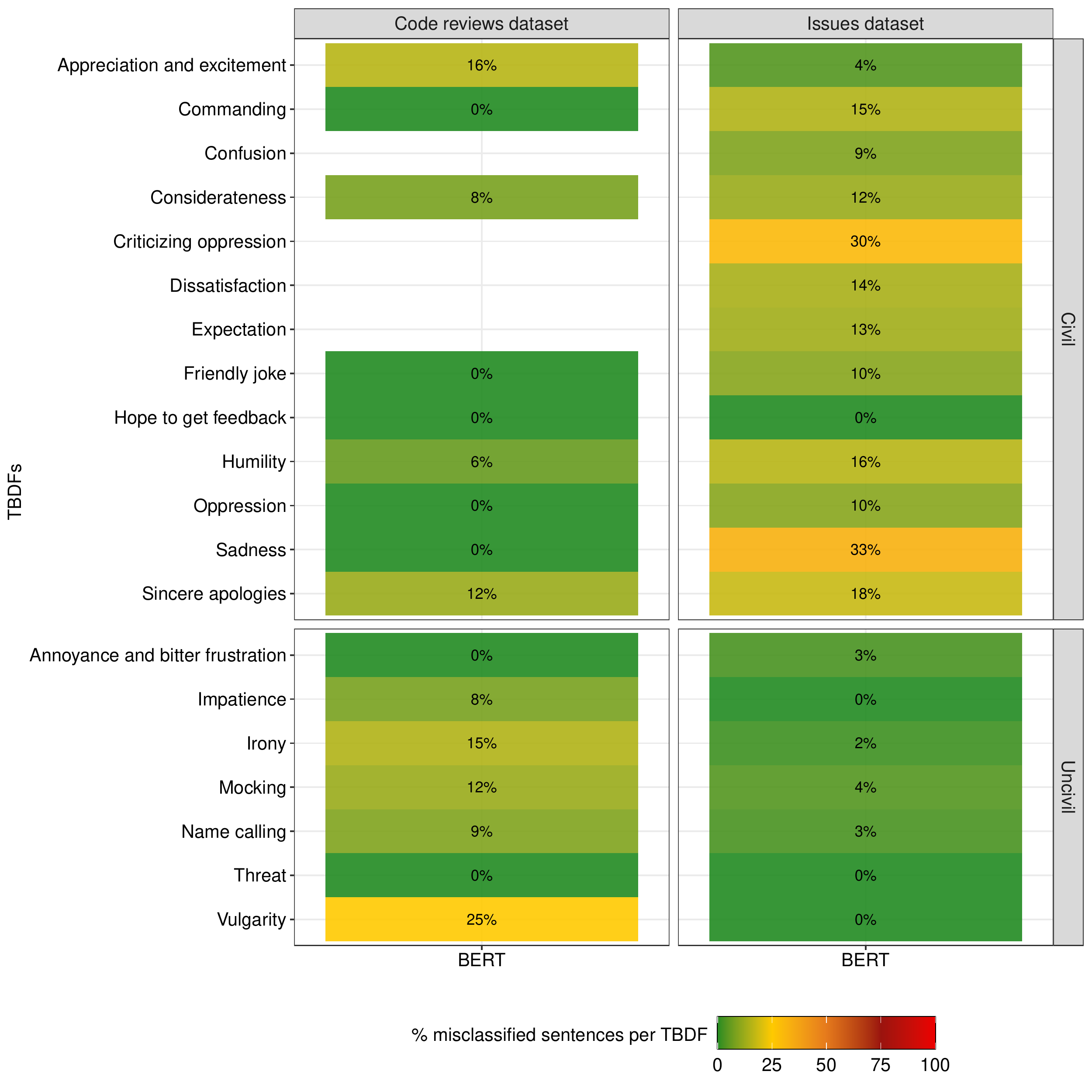}
\caption{\CHANGED{Percentage of misclassified sentences per TBDF for BERT considering the previous email or comment.}}
\label{fig:discussion_rq2}
\end{figure}

\begin{mdframed}[backgroundcolor=black!10,rightline=true,leftline=true]
\textbf{Summary RQ4 (part 2):} When the previous email or comment is added, even though the accuracy of detecting some TBDFs (e.g., \textit{friendly joke}) by BERT has improved, the accuracy of many other TBDFs has deteriorated.
\end{mdframed}

\subsubsection{Misclassified TBDFs in cross-platform settings}

\textbf{When training on code reviews and testing on issues, BERT misclassifies more than 50\% of the sentences demonstrating \textit{confusion}, \textit{dissatisfaction}, \textit{oppression}, \textit{criticizing oppression}, \textit{commanding}, \textit{considerateness}, and \textit{sadness}} (see Figure~\ref{fig:discussion_rq3}). It is expected that BERT's performance is decreased for \textit{confusion}, \textit{criticizing oppression}, \textit{dissatisfaction}, and \textit{expectation}, since those TBDFs are not present in the code reviews dataset; hence, BERT never saw examples of these TBDFs in the training set. Interestingly, in this setting, BERT classifies all instances of \textit{hope to get feedback} correctly, and it misses up to 16\%; \textit{name calling} (7.14\%), \textit{annoyance and bitter frustration} (11.09\%), \textit{impatience} (13.58\%), and \textit{sincere apologies} (16.13\%). \textbf{The classical machine learning models tend to misclassify more sentences than BERT, except for the \textit{vulgarity} with KNN, and \textit{irony} and \textit{mocking} with LR.}

\textbf{When training on issues and testing on code reviews, BERT tends to misclassify more than 50\% of the sentences classified as \textit{sadness}, \textit{friendly joke}, and \textit{considerateness}.} Similar to the other cross-platform setting, BERT classifies all instances of \textit{hope to get feedback} correctly and it misses only 9.23\% of sentences related to \textit{vulgarity}, 11.88\% \textit{mocking}, and 12.36\% \textit{name calling}. Interestingly, in this setting \textbf{classical machine learning models are better than BERT to classify various TBDFs}, such as \textit{commanding} (CART and LR), \textit{friendly joke} (LR), \textit{humility} (LR), \textit{oppression} (LR), \textit{sincere apologies} (LR), \textit{annoyance and bitter frustration} (RF), \textit{impatience} (RF), \textit{irony} (CART, KNN, RF, and SVM), \textit{mocking} (RF and SVM), \textit{name calling} (SVM), and \textit{threat} (NB and SVM).

\begin{figure}[t]
\centering
\includegraphics[clip, width=\linewidth]{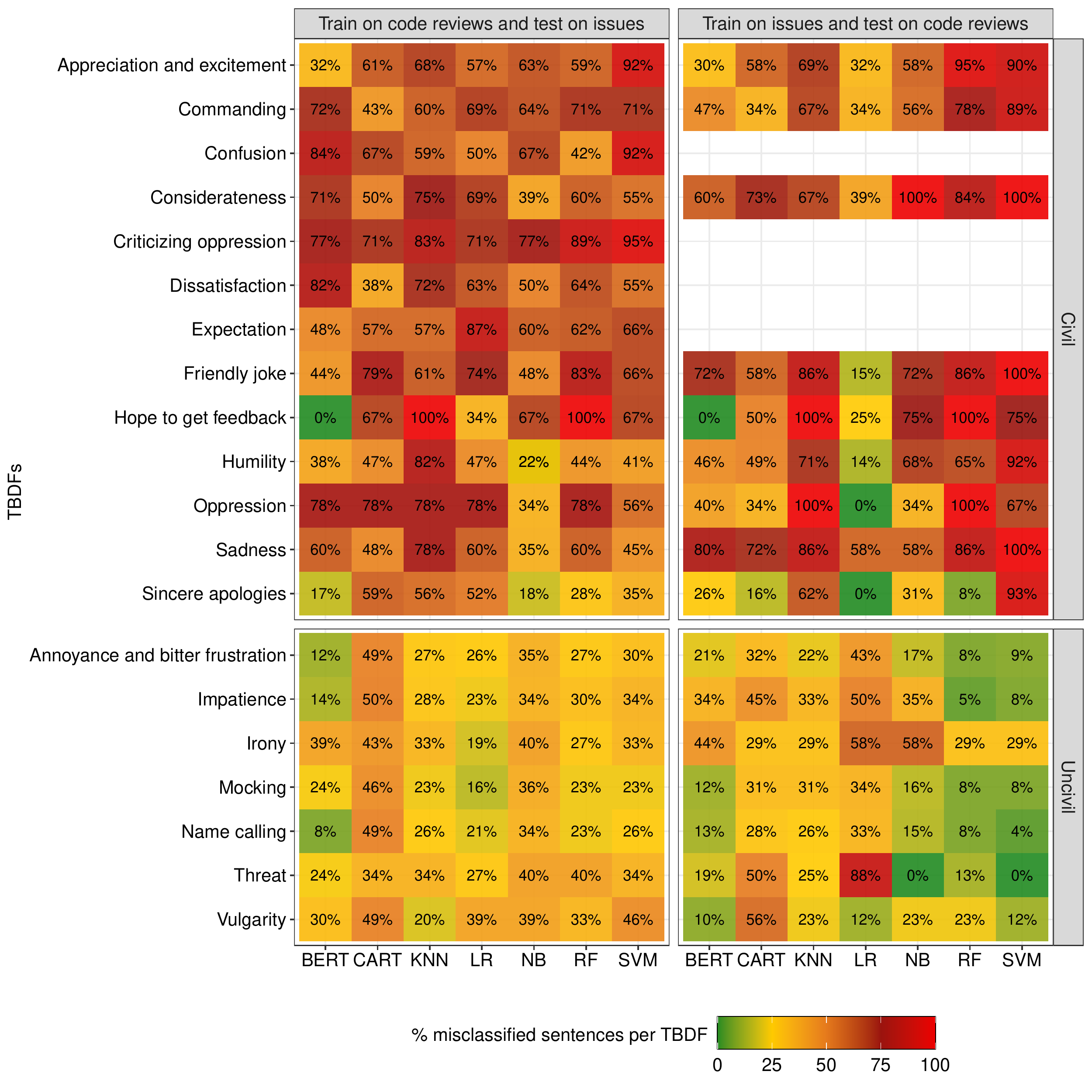}
\caption{\CHANGED{Percentage of misclassified sentences per TBDF in cross-platform settings.}}
\label{fig:discussion_rq3}
\end{figure}

\begin{mdframed}[backgroundcolor=black!10,rightline=true,leftline=true]
\textbf{Summary RQ4 (part 3):} In a cross-platform setting, the accuracy of all TBDFs degraded for BERT. The TBDFs that are the most challenging to correctly classify are \textit{commanding}, \textit{considerateness}, \textit{oppression}, and \textit{sadness}.
\end{mdframed}}

\section{Discussion}
\CHANGED{\subsection{Synthesizing the key results}}
Our results show that \textbf{BERT performs better than classical machine learning models} in both non-tone-bearing/tone-bearing and civil/uncivil classification on code review emails and issue comments, with a F1-score higher than 0.9. This result is similar to the ones found in the literature for the classification of sentiments~\cite{biswas2020achieving, batra2021bert, wu2021bert} and offensive language~\cite{cheriyan2021towards} in different software engineering artifacts (such as Stack Overflow posts, GitHub issues, API reviews, Jira issues, Gerrit code reviews, Gitter, and Slack). Hence, this paper contributes to the literature by demonstrating that BERT can also be used to classify incivility in code review emails and issue discussions. Furthermore, our results demonstrate that \textbf{classical machine learning techniques tend to underperform when classifying tone-bearing code review emails and issue comments and civil sentences in both datasets}. Since BERT has a F1-score greater than 0.90 when identifying both of these target classes, it is unclear what are the cases that classical machine learning models miss and that BERT does not. 

However, we found that \textbf{adding the previous code review email and issue comment makes the prediction of non-tone-bearing/tone-bearing code review emails and issue comments and civil/uncivil sentences worse, if not unchanged}. This result echoes Murgia \etal~\cite{murgia2014developers}, which demonstrated that the previous discussion does not help when classifying emotions in issue comments. But at the same time, this result is counterintuitive to us, since based on our experience of manually classifying incivility in code reviews and issue discussions~\cite{ferreira2021shut, ferreira2022how}, we would expect that adding some context would improve the classifiers' performance. One explanation for this is that such conversations are not ``flat'' or ``linear''; \ie the context is more complex than the immediate previous email. We plan to examine ways to capture this complexity in future work.

Finally, we found that \textbf{the classifiers' performance degraded in a cross-platform setting}, with BERT still being the best-performing model with F1 and nMCC scores below 0.7. Similarly, Qiu \etal~\cite{qiu2022detecting} found that classifiers' performance degraded when training on toxic issues and code review comments and testing on pushback in code reviews and vice versa. While performance degradation is expected, BERT's classification is still way better than random, especially for the non-tone-bearing and uncivil classes. However, whether this performance is satisfactory in practice needs future investigation.

\CHANGED{\subsection{Comparison with toxicity detectors}
While civility and toxicity are different concepts, we examined two state-of-the-art toxicity detectors on our datasets to see their performance: (1) an SVM-based model STRUDEL~\cite{ramanstress} and (2) a BERT-based model ToxiCR~\cite{sarker2023automated}. Particularly, we retrained and evaluated the two pipelines with our dataset for both non-tone-bearing/tone-bearing and civil/uncivil classification tasks (CT1 and CT2, respectively). We found that STRUDEL achieved a macro-average F1 of 0.62 for CT1 and 0.72 for CT2 for the code review dataset, and 0.67 for CT1 and 0.68 for CT2 for the issue discussion dataset. This performance is better than our SVM model but is inferior than our BERT pipeline. Moreover, the BERT-based ToxiCR resulted in a macro-average F1 of 0.57 (CT1) and 0.68 (CT2) for the code review dataset, and 0.64 (CT1) and 0.73 (CT2) for the issue discussion dataset. This performance is inferior than our BERT pipeline. These results indicate that civility detection is a unique problem and our pipeline that incorporate data augmentation and class balancing can help achieve better results.}

\subsection{Lessons learned}

Based on our results, we provide insights into the lessons learned for using automated incivility detection techniques.

\begin{mdframed}[backgroundcolor=black!10,rightline=true,leftline=true]
\textbf{Lesson 1:} \CHANGED{Adding the previous email/comment} helps to accurately classify challenging TBDFs. However, this approach should not be used to classify more straightforward TBDFs.
\end{mdframed}

Our results suggest that some TBDFs are more challenging to be classified than others. For such cases, \CHANGED{adding the previous email/comment} is a way to mitigate the problem. For example, in the code review dataset, \textit{friendly joke}, \textit{commanding}, \textit{sadness}, and \textit{threat} can only be accurately classified with \CHANGED{the previous email/comment}. Other TBDFs seem to be more straightforward to be classified, such as \textit{appreciation and excitement, sincere apologies, impatience, and mocking}. \textit{Irony} and \textit{vulgarity} are still challenging to be classified and may require TBDF-specific classifiers.

\begin{mdframed}[backgroundcolor=black!10,rightline=true,leftline=true]
\textbf{Lesson 2:} If a GPU is not available to train BERT, Random Forest can be used to precisely classify the code review dataset and SVM can be used for the issues dataset.
\end{mdframed}

We found that BERT is the best model to identify incivility in open source discussions with high precision and high recall. However, BERT is a computationally expensive algorithm that requires a GPU. If a GPU is not available, for the code review dataset, the Random Forest model is the best option being able to precisely classify 75\% of the data (although it will miss 48\% of the cases). For the issues dataset, SVM is a good compromise with a precision of 0.93. Depending on the scenario, recall might be more important than precision. In that case, the Logistic Regression can be used for the code review dataset ($recall = 0.68$) and the issues dataset ($recall = 0.75$).

\begin{mdframed}[backgroundcolor=black!10,rightline=true,leftline=true]
\textbf{Lesson 3:} It is feasible to use BERT to classify non-tone-bearing and uncivil code review emails and issue comments in a cross-platform setting when a manually annotated gold standard is not available.
\end{mdframed}

Our results show that, in a cross-platform setting, BERT can be used to classify non-tone bearing discussions with a precision and recall greater than 0.90. For the uncivil class, BERT's precision and recall are greater than 0.75. Thus, it can be used in certain cross-platform detection use cases.

\section{Threats to Validity}

We discuss threats to the study validity~\cite{wohlin2012experimentation} as follows.

\paragraph{\textbf{Construct Validity}} 
The incivility dataset from Ferreira \etal~\cite{ferreira2021shut, ferreira2022how} might contain noise (such as the source code, words other than English, special characters, \etc.) that can affect the models' performance. To mitigate this threat, we followed strict steps to preprocess the text (Section~\ref{sec:data_preprocessing}). Hence, we expect to have removed the noise in the data. Furthermore, the set of features used for the classical models might not represent all confounding factors in incivility. We minimize this threat by adopting the features from a previous work focused on characterizing sentences in issue discussions~\cite{arya2019analysis}. Finally, when assessing if the contextual information helps to detect incivility, the presence of civil or uncivil words and non-tone-bearing or tone-bearing words in the previous code review email/issue comment can affect the models' performance. To mitigate this threat, we computed the number of previous emails and comments that are \textit{non-tone-bearing} $\to$ \textit{tone-bearing} and \textit{civil} $\to$ \textit{uncivil} and vice versa. For CT1, we found up to 7.76\% code review emails and up to 8.05\% issues comments in this situation. For CT2, we found up to 5.36\% code review sentences and up to 12.68\% of issue sentences in this situation. Given the relatively low number of datapoints in such a situation, we think that our results will not be highly affected by that.




\paragraph{\textbf{Internal Validity}}

The models might overfit due to the small number of labeled datapoints. To address this problem, we implemented four data augmentation techniques with eight combinations of hyperparameters to ensure optimal results. Additionally, the imbalance in the datasets may lead to poor performance. To minimize this threat, we compared three class balancing techniques and assessed their performance. Finally, the choice of hyperparameters might affect the results. For that, we did hyperparameter optimization on all seven models using search spaces defined in the literature. Additionally, our model evaluation is based on the average metric values of a 5-fold cross-validation.


\paragraph{\textbf{Conclusion Validity}}

All our validations (either to find the best hyperparameter or to compare the models) are based on the nMCC metric, which is known to be more interpretable and to have more robust results than other performance metrics.

\paragraph{\textbf{External Validity}}

Our incivility classifiers are limited to code reviews of rejected patches of the Linux Kernel Mailing List and GitHub issues locked as too heated. Hence, our results may not be generalizable to other software engineering communication artifacts; this includes the results of cross-platform performance. Concerning the features used by the classical techniques, we have experience with incivility studies and we manually coded the data in our previous work~\cite{ferreira2021shut, ferreira2022how}. Hence, we were able to assess if the features are accurate to the incivility domain. Finally, our results are confined to the models implemented in this study. It is unknown if other models that would perform better for incivility classification.
\section{Conclusion}

Open source communities have developed mechanisms for handling uncivil discourse. However, the current mechanisms require considerable manual effort and human intervention. Automated techniques to detect uncivil conversations in open source discussions can help alleviate such challenges. In this paper, we compared six classical machine learning techniques with BERT when detecting incivility in open source code review and issue discussions. Furthermore, we assessed if adding \CHANGED{the previous email/comment} improves the classifiers' performance and if the seven classifiers can be used in a cross-platform setting to detect incivility. In our analysis, we identified BERT as the best model to detect incivility in both code review and issue discussions. Furthermore, our results show that classical machine learning models tend to underperform when classifying tone-bearing and civil conversations. We also found that \CHANGED{adding the previous email/comment} does not improve BERT's performance and that the classifiers' performance degraded in a cross-platform setting. Finally, we provide three insights on the discussion features that the classifiers misclassify when detecting incivility. These insights will help future work that aims at leveraging discussion features in automated incivility detection applications, as well as improving cross-platform incivility detection performance.

\section*{Acknowledgements}
The authors would like to thank Calcul Québec for the computing hardware that enabled them to run the experiments of this study. The authors also thank the Natural Sciences and Engineering Research Council of Canada for funding this research through the Discovery Grants Program [RGPIN-2018-04470].





\bibliographystyle{elsarticle-num} 
\bibliography{bibliography.bib}





\end{document}